\documentclass[12pt,prd,nofootinbib,superscriptaddress,tightenlines,showpacs]{revtex4-1}
\usepackage{amsmath}
\usepackage{amssymb}
\usepackage{graphicx}

\begin{document}

\title{Black hole binary inspiral and  trajectory dominance}

\author{Richard H.~Price} \affiliation{Department of Physics \&
  Astronomy, and Center for Advanced Radio Astronomy,
University of Texas at Brownsville, Brownsville
  TX 78520}

\author{Gaurav Khanna} \affiliation{Department of Physics, University
  of Massachusetts, Dartmouth, MA 02747}

\author{Scott A.~Hughes} \affiliation{Department of Physics and MIT
  Kavli Institue, MIT, 77 Massachusetts Ave., Cambridge, MA 02139}

\affiliation{Canadian Institute for Theoretical Astrophysics,
  University of Toronto, 60 St.\ George St., Toronto, ON M5S 3H8,
  Canada}
\affiliation{Perimeter Institute for Theoretical Physics, Waterloo, ON
  N2L 2Y5, Canada}

\begin{abstract}
Gravitational waves emitted during the inspiral, plunge and merger of
a black hole binary carry linear momentum. This results in an
astrophysically important recoil  to the final merged black hole, a ``kick'' that
can eject it from the nucleus of a galaxy.  In a previous paper we
showed that the puzzling partial cancellation of an early kick by a late
antikick, and the dependence of the cancellation on black hole spin, can be understood
from the phenomenology of the linear momentum waveforms. Here we connect that
phenomenology to its underlying cause, the spin-dependence of the
inspiral trajectories. This insight suggests that the details of
plunge can be understood more broadly with a focus on inspiral
trajectories.
\end{abstract}

\maketitle

\section{Introduction}\label{sec:intro}

During the inspiral and merger of an asymmetric black hole (BH) binary,
the linear momentum that is emitted results in a reaction, a ``kick,''
to the final merged black hole. This kick can be strong enough to
eject the merged final black hole from its host active galactic
nucleus.  See, for example,
Refs.\ \cite{vgm10,zrdzp10,gmmc11,ssh11,bclh11} for recent work
discussing astrophysical implications of black hole kicks.
Observational confirmations of the predicted ``runaway'' black holes
are now starting\cite{runawayAGNs}.

Theoretical predictions of kicks have been based largely on supercomputer numerical
computations  of the nonlinear equations of general
relativity for black hole inspiral and merger. These codes
are now capable of evolving almost any initial binary
configuration. Explorations and good guesses have been made that have
led to ``superkick'' configurations with very large ejection
velocities of the final hole\cite{bigkicks}. What is missing is a 
picture of the process simple enough 
so that physical insights can be used, as they usually are in
physics. This has been a main motivation for the visualization project
by the Caltech-Cornell group\cite{visualization} in which ``tendex and
vortex'' lines are used for visualization of the relativistic
gravitational fields.  

Here we give a simple and compelling picture of the generation of at least some 
aspect of kicks, a picture based on the idea that in binary inspiral main features
of emission are to be understood as manifestations of the details of trajectories.
What is perhaps most important about the success of this picture is that it
suggests that ``trajectory dominance'' may be a key to a phenomenological 
understanding of binary inspiral emission more generally.

The remainder of this paper is organized as follows. In
Sec.~\ref{sec:review} we briefly review the spin-dependent
kick-antikick cancellation for equatorial orbits, along with our
phenomenological explanation of the cancellation and its spin
dependence.  Section ~\ref{sec:orbits} then looks at inspiral
orbits. It is shown that the qualitative characteristics of these orbits
correlate with black hole spin in a way that suggests that it is the
orbital shapes that explain the different characteristics of linear
momentum emission for prograde vs.~retrograde orbits, and for different
spins. In this section it is also shown that the root of the different
orbital characteristics (and hence of the kick correlation with spin
and orbital direction) is the relationship of particle orbital
angular momentum and angular velocity in the spacetime of a rotating
hole.
Section~\ref{sec:tests} then ``tests'' the hypothesis of trajectory dominance
with two classes of numerical experiments.  In the first, it is shown that 
a Kerr particle trajectory placed in a Schwarzschild spacetime gives substantially
the same gravitational wave emission as it does in the Kerr spacetime for which it is a geodesic.
The second class of tests is limited to retrograde orbits in Kerr spacetimes. It is 
shown that the burst of radiation from retrograde orbits arises from the 
reversal of angular velocity of the inspiral trajectory.
We discuss the
implications of these results in Sec.~\ref{sec:conc}.

\section{Phenomenological explanation of the Kick-antikick cancellation for quasicircular equatorial 
orbits}\label{sec:review}

During the BH inspiral-plunge-merger (IPM) the
gravitational wave (GW) emission carries away linear momentum, and a
net linear momentum emission builds up in some direction.  A strange
attribute of the linear momentum was noted by Schnittman {\it et al.}\cite{SchnittmanEtAl}
in their computational studies of the IPM of comparable mass BHs, with
spin angular momentum perpendicular to the orbital plane.  The net
linear momentum in some direction would grow during the inspiral
phase then start to decrease at the plunge. For certain models the
decrease removed most of the 
momentum emitted earlier. Subsequently, Sundararajan {\it et
al.}\cite{SundararajanEtAlI} studied the phenomenon further with the
flexibility and efficiency of particle perturbation techniques. Their
results, for ``particles'' orbiting in the equatorial plane of a
spinning black hole, included models in which 97\% of the kick was
cancelled by a late term antikick.  It was noted in these studies that
the extent of cancellation is strongly correlated with black hole
spin and strongly dependent on whether the orbital motion is prograde
(orbital and spin angular momentum aligned) or retrograde
(antialigned). We shall call this puzzling cancellation, along with its dependence 
on the orbit and the BH spin, the ``cancellation phenomenon.''

This phenomenon was somewhat a paradox.
The early momentum emission comes from the nearly Newtonian gradual
inspiral, while the late emission is from the plunge and the quasinormal
ringing of the merger. 
It seemed remarkable that the early process could
somehow ``set up'' the late process to generate just the right amount of 
linear momentum so that for some models the late momentum emission almost
completely cancelled the early emission.

As is so often the case for an ``impossible'' coincidence, the
explanation turns out to be simple, at least at one level. For
prograde orbits the component of linear momentum flux in any
direction, let us say the $\dot{P}_x$ in the $x$-direction, is an
oscillating quantity. This oscillating quantity starts with negligible
amplitude in the distant past, in effect at time $t=-\infty$; it ends
with zero amplitude at $t=+\infty$, when the quasinormal ringing dies
out. Thus, as a function of time, $\dot{P}_x$ is an oscillation inside
a modulation envelope that starts and ends at zero, and is largest
around the plunge.

The net momentum $P_x$ radiated up to some time $t$ is the integral of
$\dot{P}_x$ from early time up to time $t$. The total
$P_x$ radiated for the entire IPM process,
$\int_{-\infty}^{\infty}\dot{P}_x\,dt$ is the integral of an
oscillating quantity. In that integral, the positive phases and
negative phases of the oscillation will tend to cancel.  Due to the changing
amplitude of the oscillations the cancellation will not be complete;
some net momentum can be radiated.  The more rapidly the amplitude
changes, the larger the result for the total momentum radiated. The
total momentum in fact is easily shown to be a decreasing function of
the characteristic time scale for the change in the amplitude divided
by the characteristic period of the oscillations. (For details see
Ref.~\cite{paper1}, Hereafter Paper 1.)

  \begin{figure}[h]
  \begin{center}
  \includegraphics[width=.4\textwidth ]{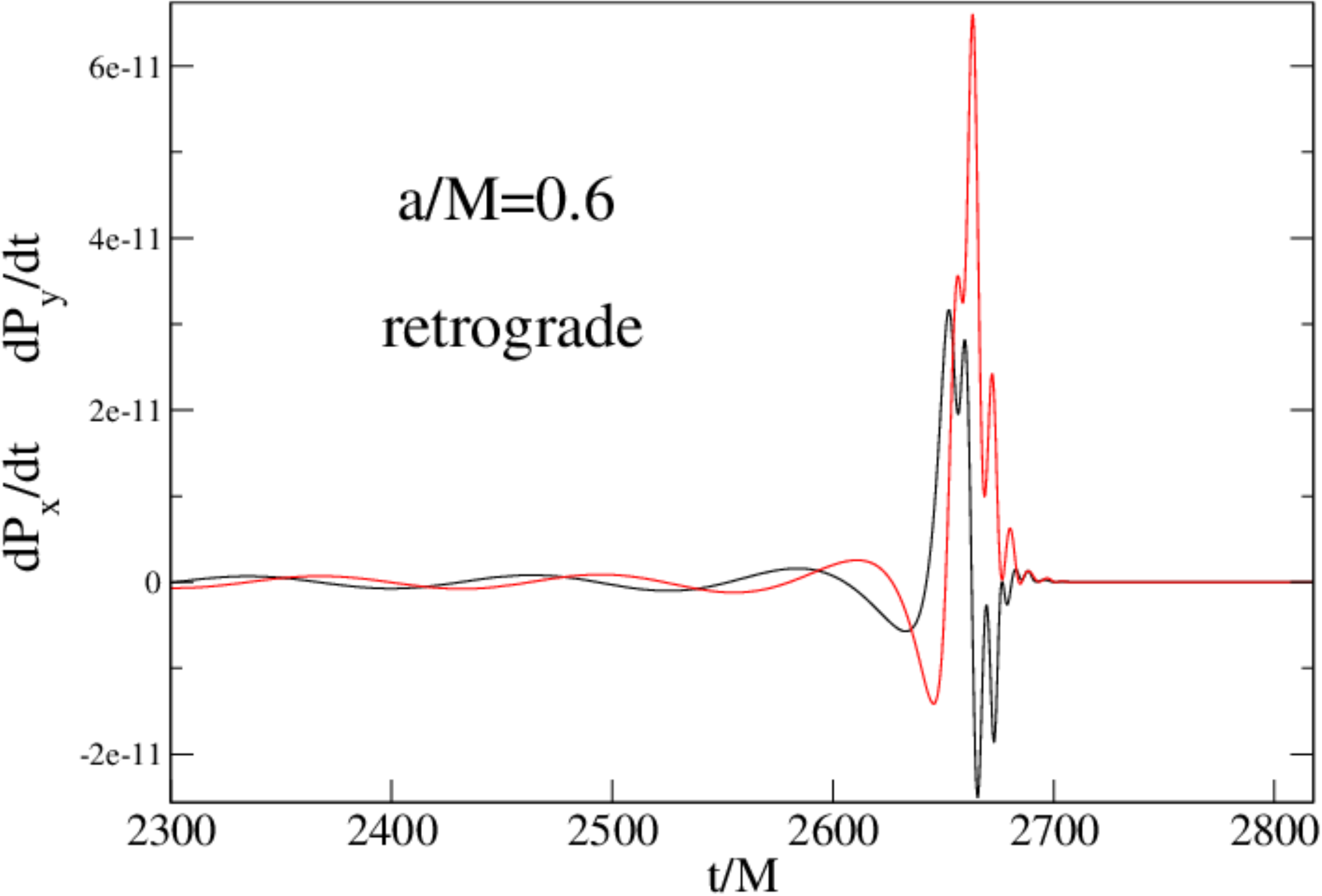}
  \includegraphics[width=.4\textwidth ]{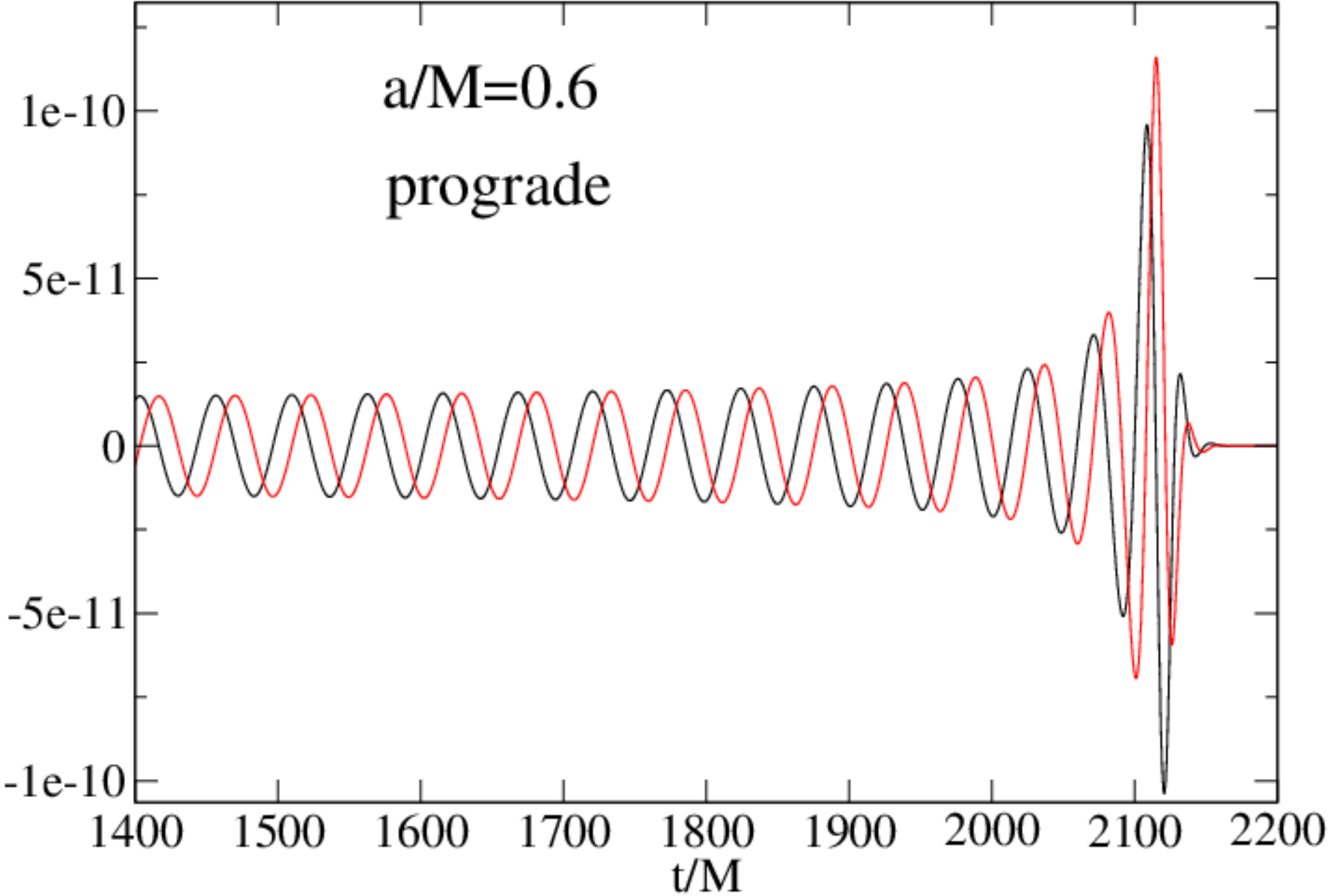}\\

\vspace{10pt}  
\includegraphics[width=.4\textwidth ]{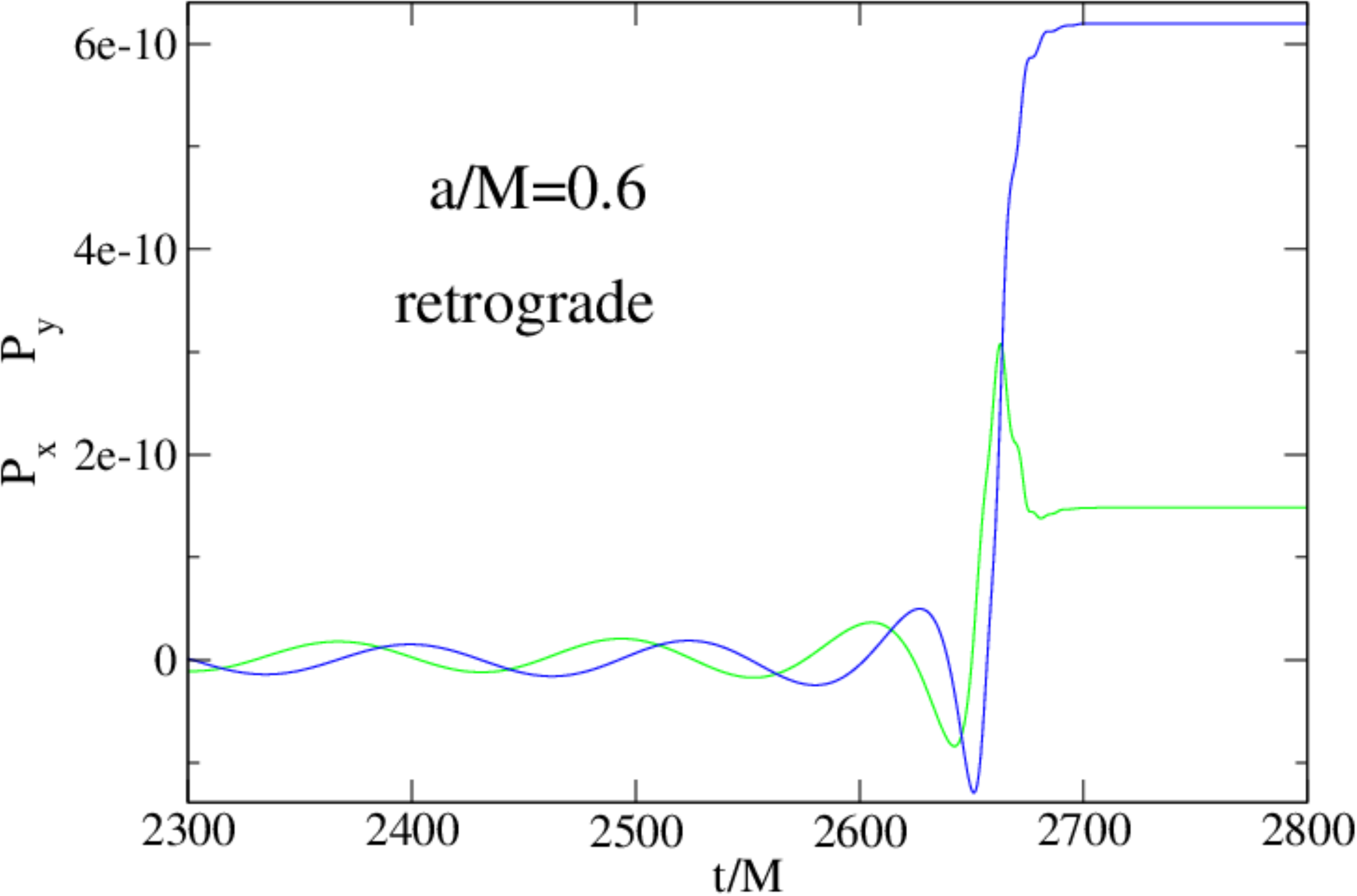}
  \includegraphics[width=.4\textwidth ]{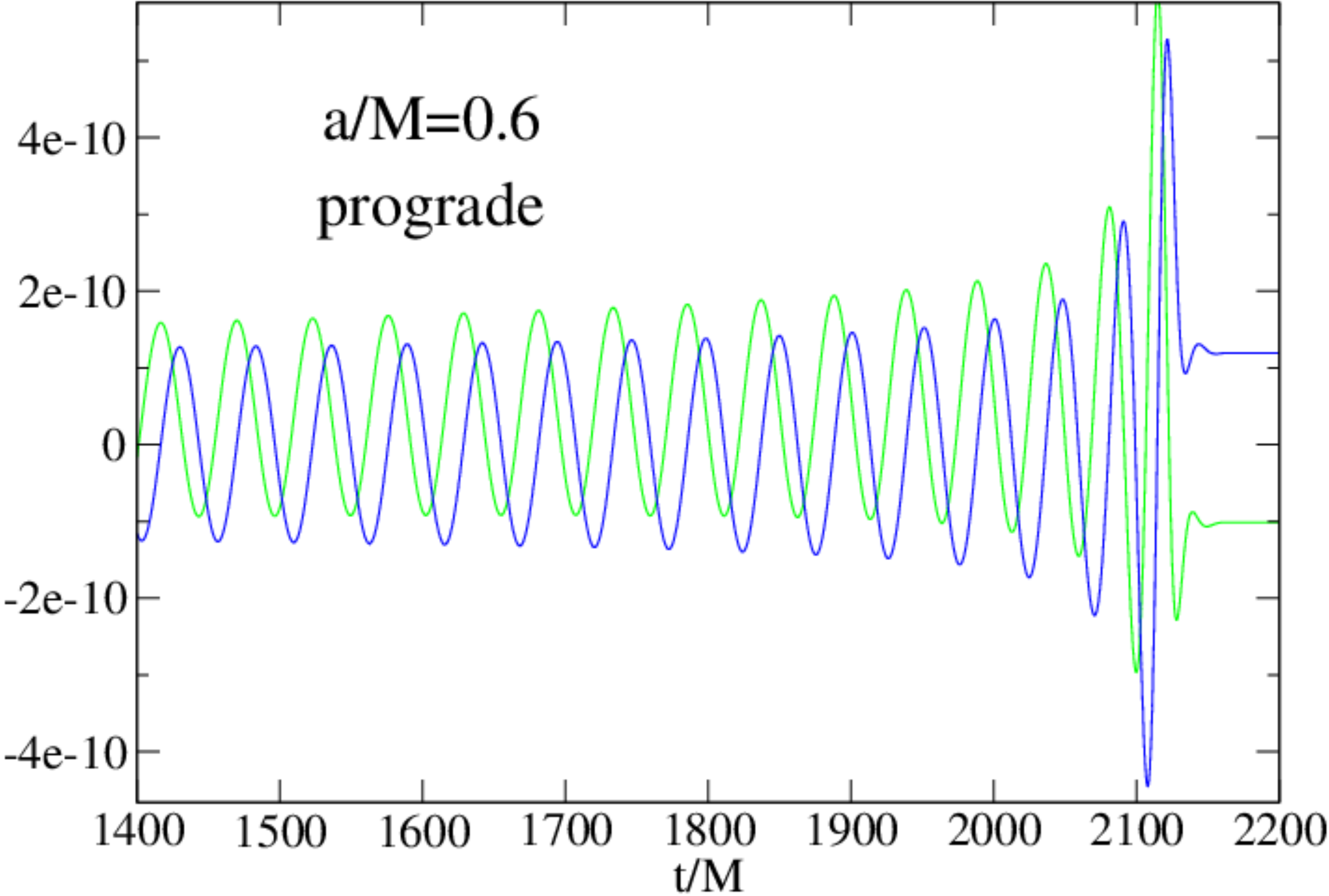}
  \caption{The top row shows, as a function of time, the momentum flux
    components $\dot{P}_x$ and $\dot{P}_y$ for both a retrograde (left
    plot) and a prograde (right) inspiral into $a/M=0.6$ spinning
    hole. The bottom row shows the components  $P_x,P_y$ of the total linear
    momentum radiated from $t=-\infty$.  }
  \label{fig:pandpdot06}
  \end{center}
  \end{figure}

  \begin{figure}[h]
  \begin{center}
  \includegraphics[width=.4\textwidth ]{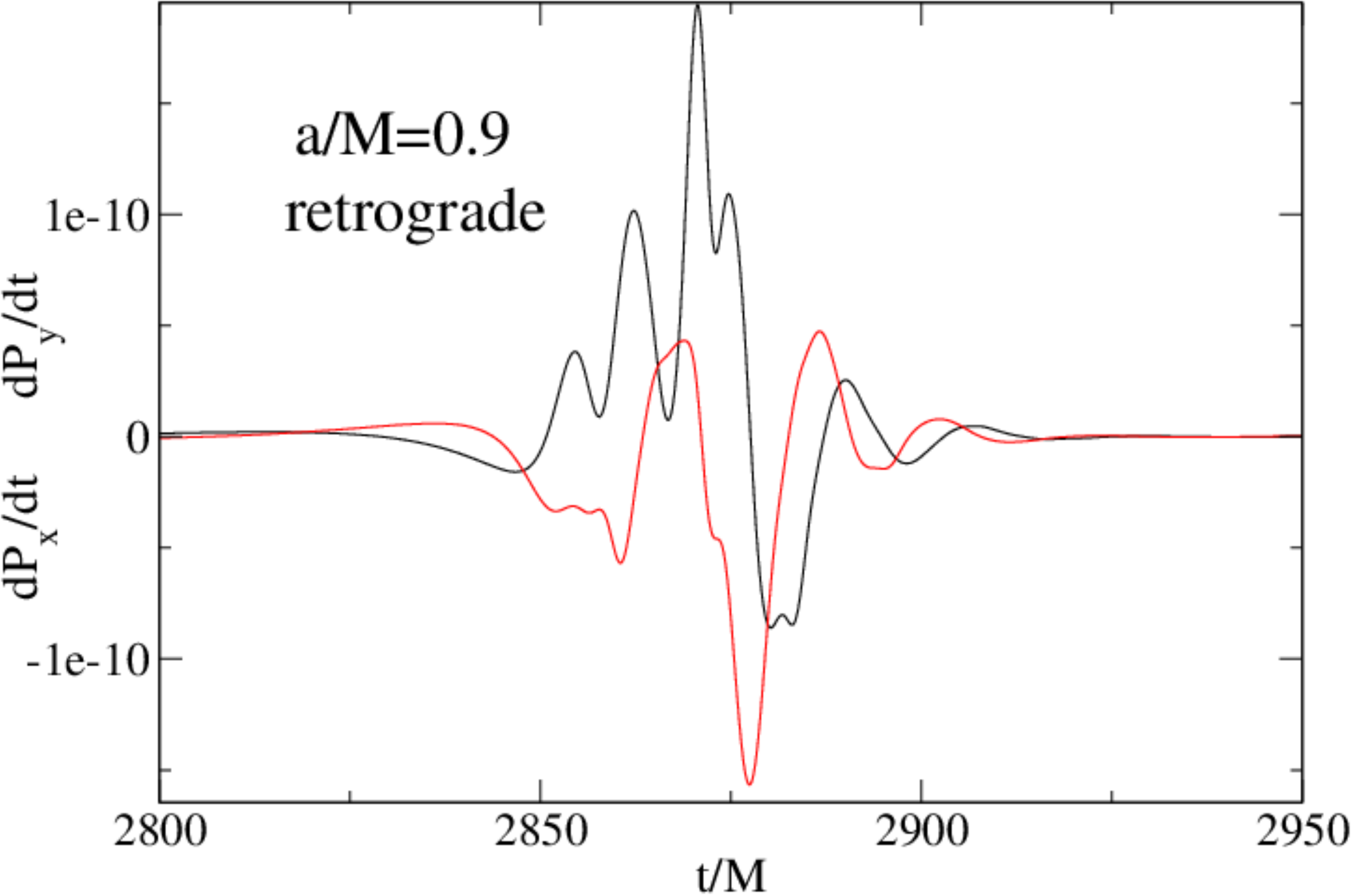}
  \includegraphics[width=.4\textwidth ]{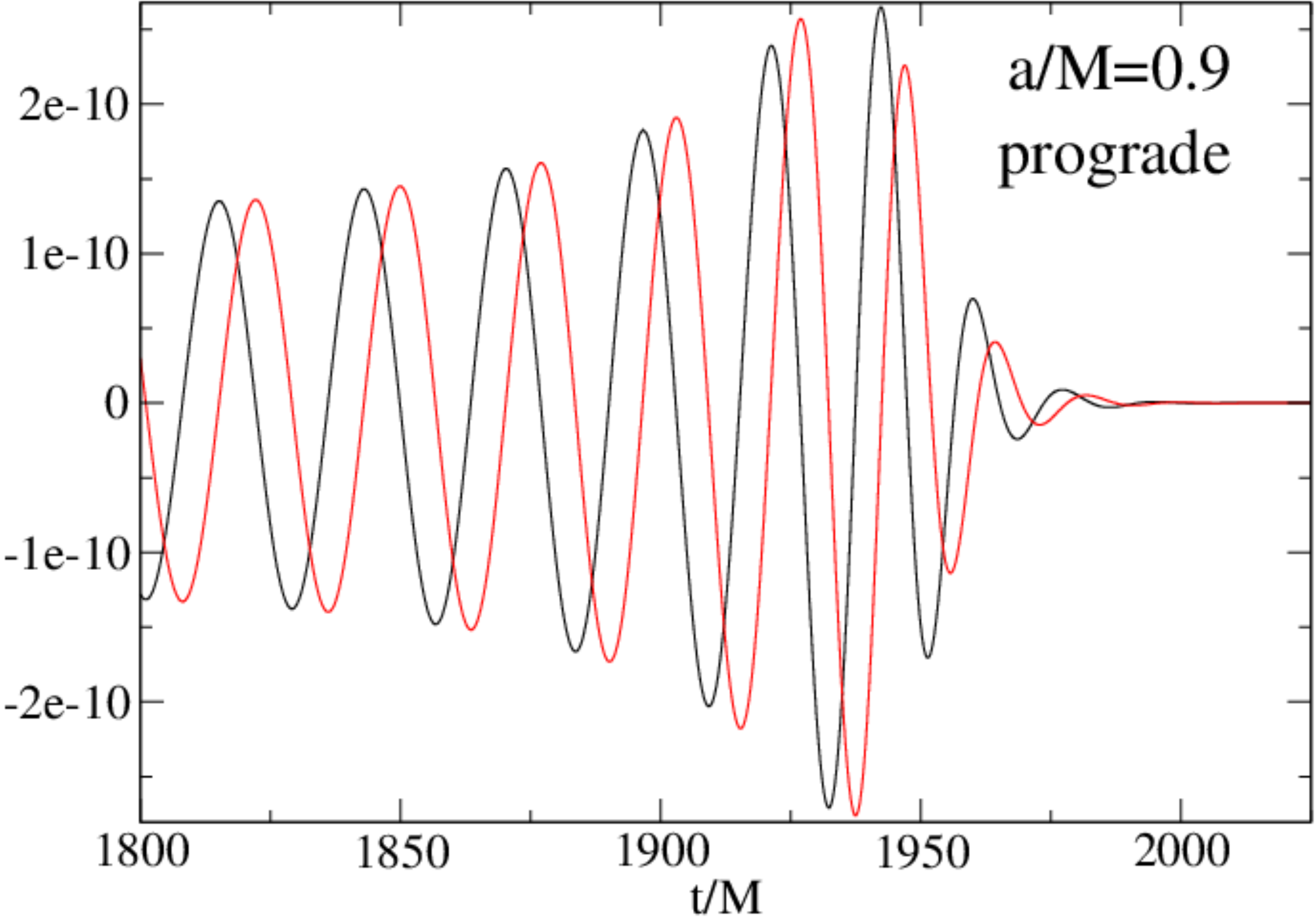}\\
  
\vspace{10pt}
\includegraphics[width=.4\textwidth ]{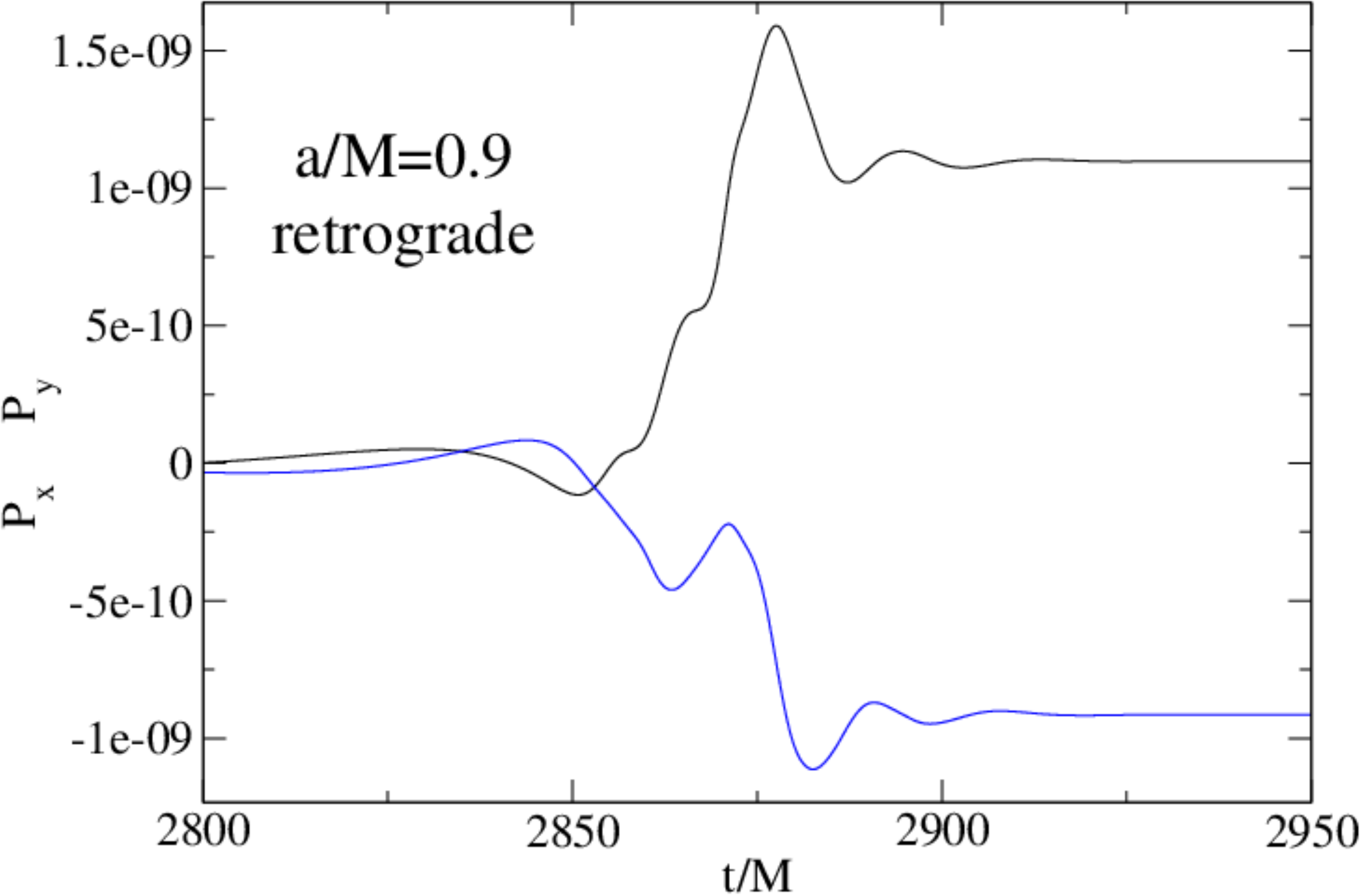}
  \includegraphics[width=.4\textwidth ]{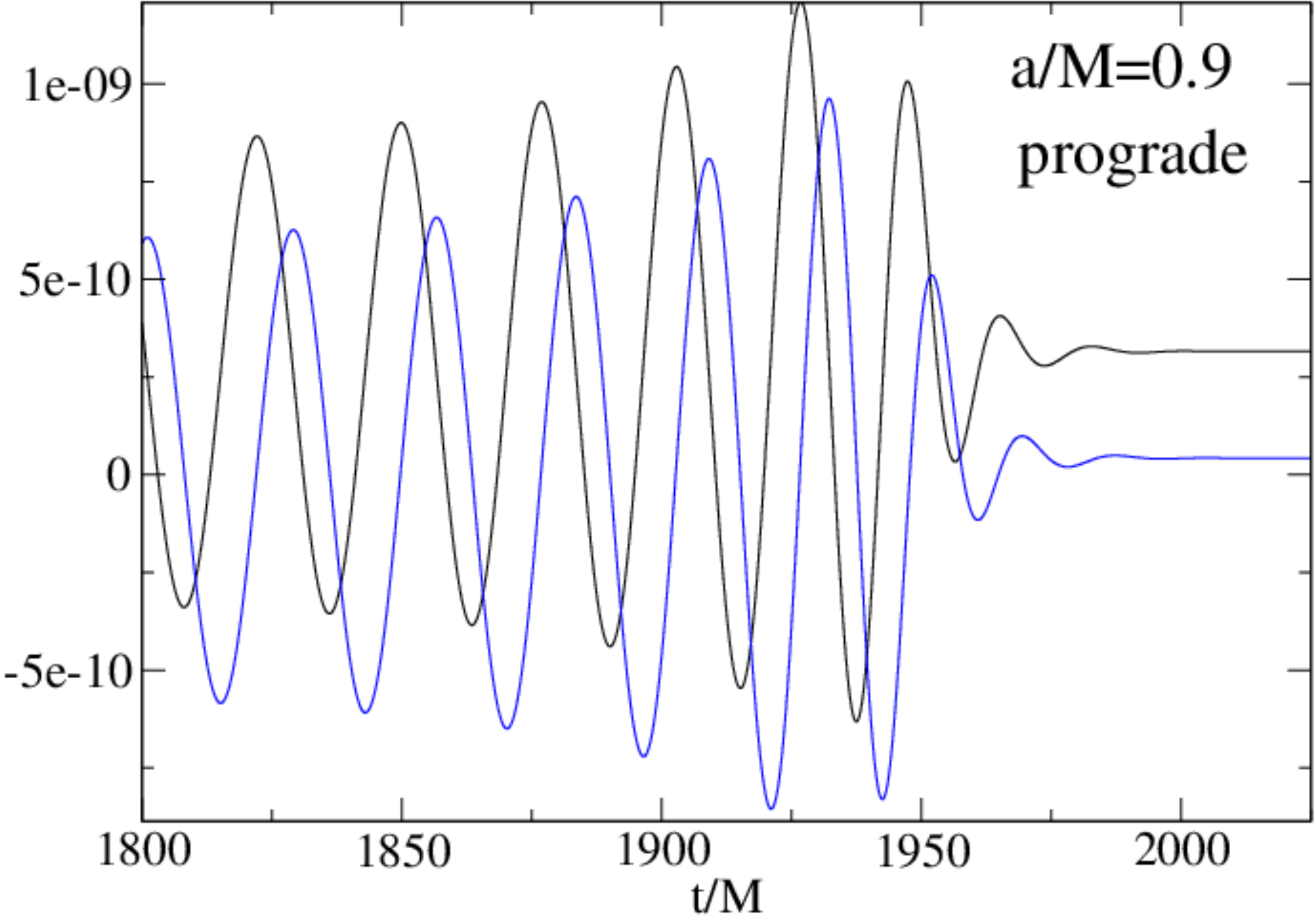}
  \caption{The same quantities as in Fig.~\ref{fig:pandpdot06} but here for 
retrograde and prograde orbits into a black hole with 
$a/M=0.9$.}
  \label{fig:pandpdot09}
  \end{center}
  \end{figure}

  \begin{figure}[h]
  \begin{center}
  \includegraphics[width=.4\textwidth ]{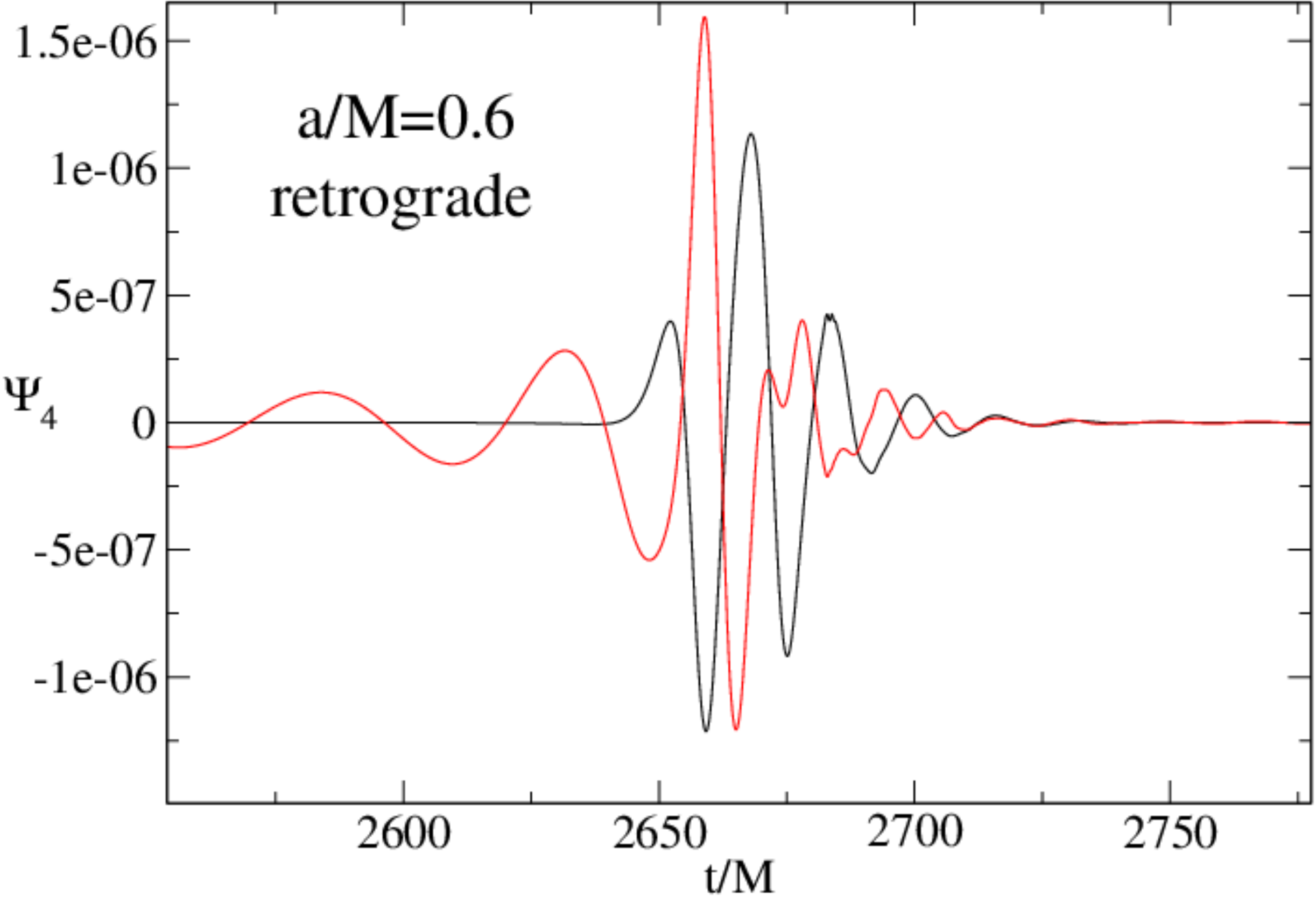}
  \includegraphics[width=.4\textwidth ]{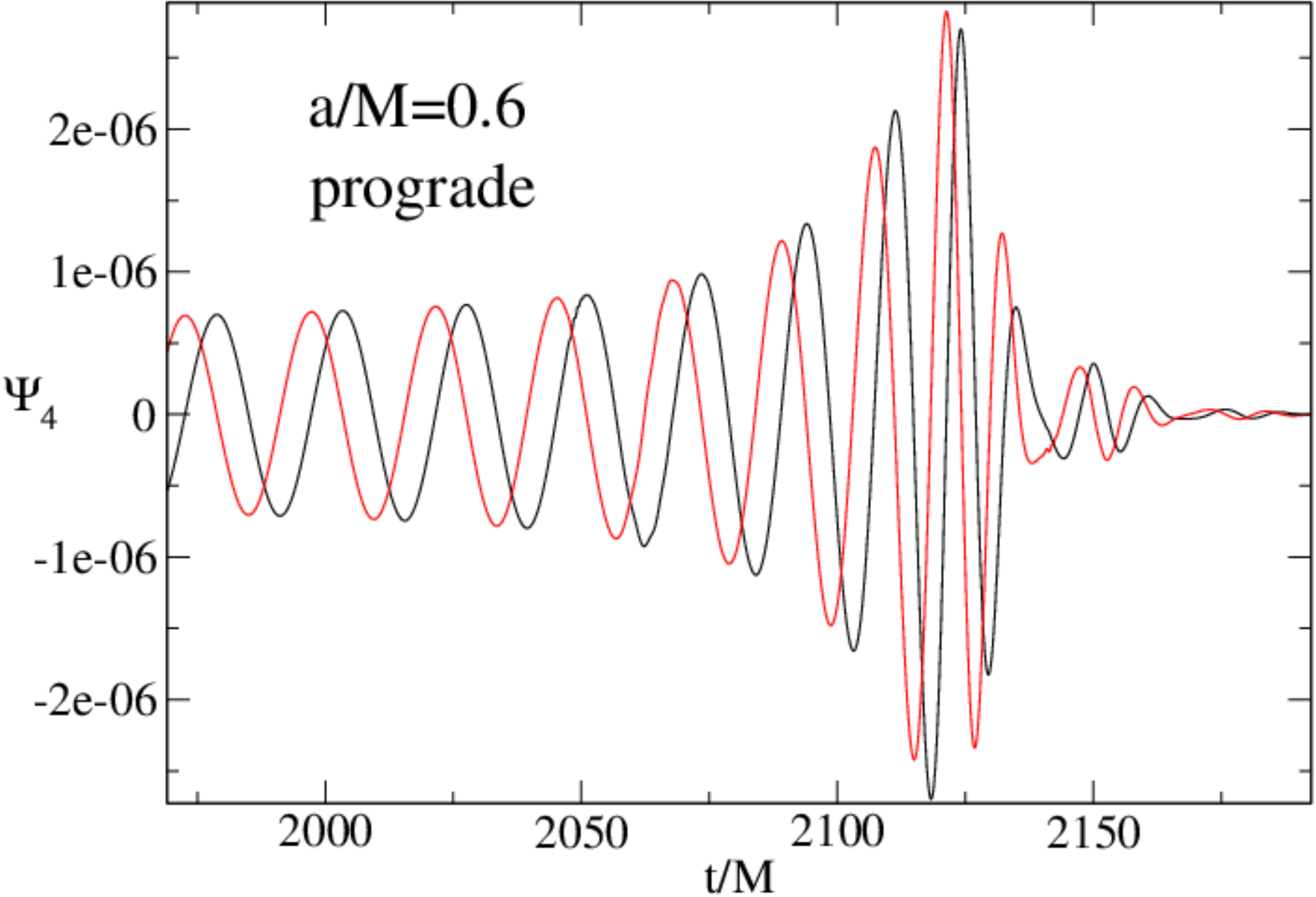}\\

\vspace{10pt}  
\includegraphics[width=.4\textwidth ]{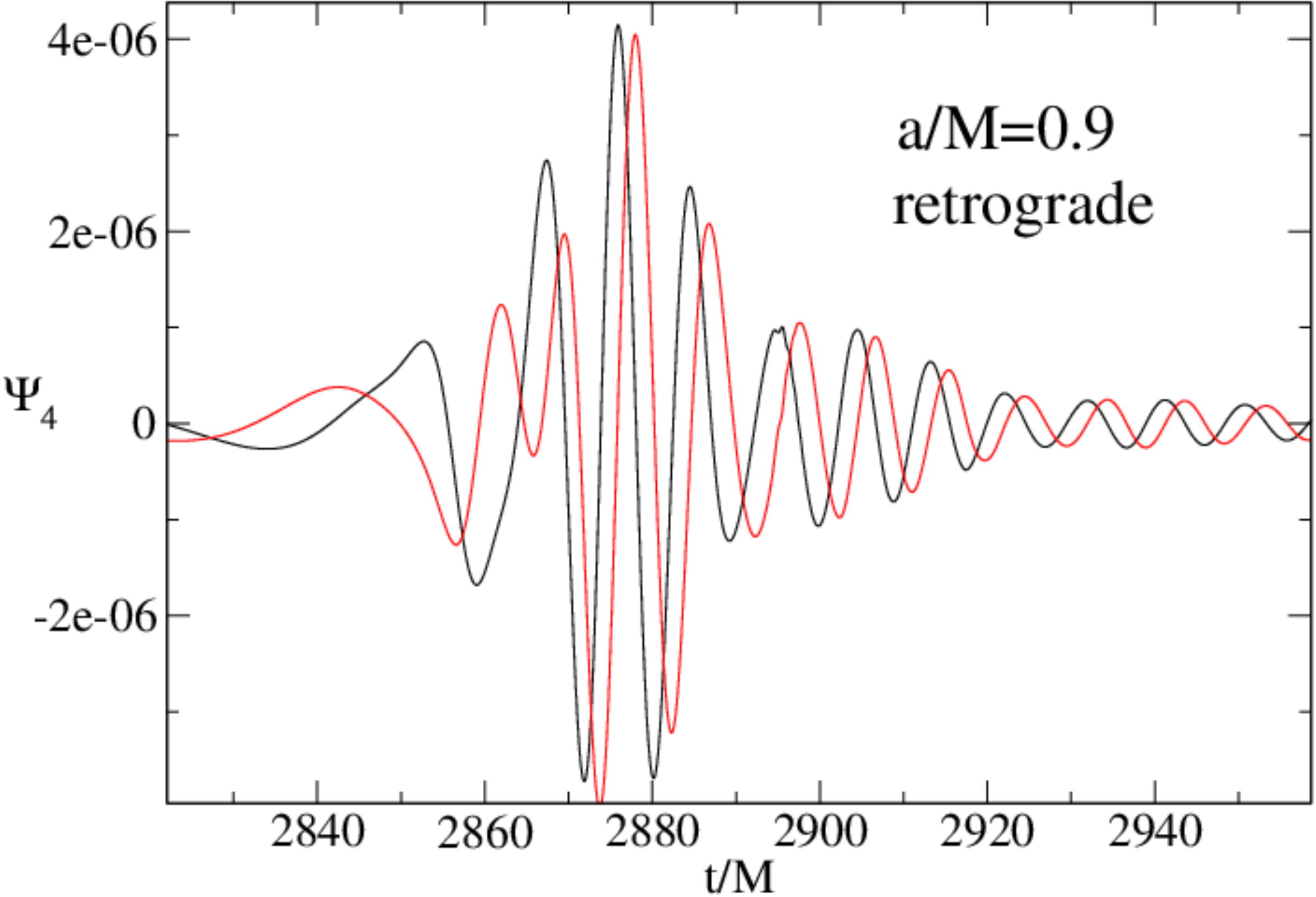}
  \includegraphics[width=.4\textwidth ]{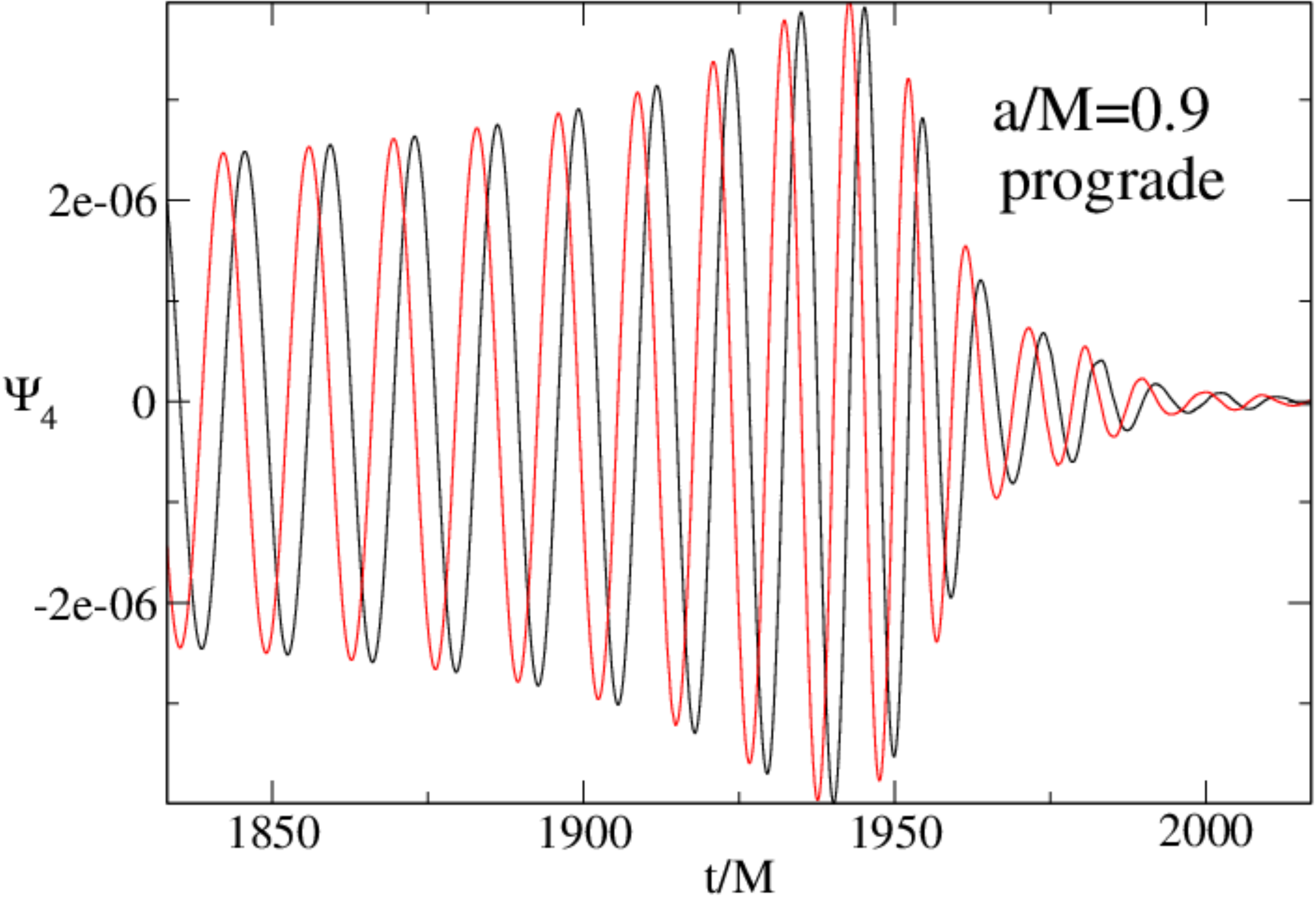}
  \caption{Waveforms, i.e., the real and imaginary part of the $m=2$ component
of the Teukolsky function $\Psi_4$ for prograde and retrograde orbits into 
Kerr holes with $a/M=0.6$ and 0.9.
}
  \label{fig:wave2s}
  \end{center}
  \end{figure}

For a very slowly varying amplitude, the components of net momentum radiated (and
hence of the net kick) $\int_{-\infty}^{\infty}\dot{P}_k\,dt$ must be
very small. Any net momentum radiated in the early increasing
amplitude part of the process, must be cancelled in the later part.
This is not a consequence of any feature of curved spacetime, but of
simple mathematics. The phenomenological explanation of the
cancellation phenomenon fits the results of the computations both for
comparable mass BHs and for EMRIs; the more gradually the amplitude
changes, the greater is the extent to which the late antikick cancels
the earlier kick. In the case of prograde equatorial orbits in EMRIs,
a more definitive statement can be made. The rate of change of the
envelope depends on the spin of the BH. Larger spin BHs show more
slowly varying amplitudes of momentum flux, and show a more nearly
complete cancellation of early and late linear momentum. Retrograde
equatorial orbits show the opposite correlation: for the most rapidly
spinning holes the linear momentum flux oscillations have the most
rapidly changing amplitude.

Figures~\ref{fig:pandpdot06} and \ref{fig:pandpdot09} illustrate the
connection between radiated linear momentum, BH spin, and direction
(prograde vs.~retrograde) for equatorial orbits.  In the top row of
each of these figures the flux of linear momentum is shown in two
arbitrary orthogonal directions $x,y$. These components are defined
from the Boyer-Lindquist coordinates $r,\phi$ by the usual flat-space
conventions $x=r\cos\phi$, $y=r\sin\phi$. In Fig.~\ref{fig:pandpdot06}
the plots on the left hand side correspond to a retrograde IPM. For
these retrograde cases the plots shows that the linear momentum
emission is largely concentrated in a burst. The net linear momentum
components (bottom row) grow suddenly upon emission of this burst and
the final linear momentum is of order of the momentum flux times the
oscillatory timescale.  The plots on the right, for a prograde orbit,
tell a very different story. Here the momentum flux is oscillatory
inside an amplitude envelope that is moderately smooth. The net
momentum emitted (lower plot) is oscillatory until the amplitude peak,
at which time a net momentum is built up, but -- unlike the retrograde
case -- this net momentum is an order of magnitude less than the
product of the momentum flux and an oscillatory time scale.  The
features shown in Fig.~\ref{fig:pandpdot06} for $a/M=0.6$ are also
present in Fig.~\ref{fig:pandpdot09} for $a/M=0.9$, but are
significantly more pronounced. For $a/M=0.9$, the jump in radiated
momentum is more sudden than for $a/M=0.6$ in the case of the
retrograde orbit, and the cancellation of the radiated momentum is
more nearly total than for $a/M=0.6$ in the case of the prograde
orbit.

In seeking an explanation for this cancellation, an important
technical question must be asked. Linear momentum cannot be generated
in a single multipole mode. Its emission therefore depends on delicate
amplitude and phase relations of different modes (in fact, the relations of even
modes with odd modes).  We must ask whether the BH spin dependence,
and the very different patterns for retrograde and prograde orbts are
the results of subtly shifting mode interactions, or whether they are
embedded more robustly in the gravitational wave emission.

This question is answered in Fig.~\ref{fig:wave2s}. Here the $m=2$
part of the Teukolsky function $\Psi_4$ is shown for retrograde and
prograde orbits both for $a/M=0.6$ and $a/M=0.9$. It is clear in
these figures that what is seen in the linear momentum flux is also
true for the gravitational waves themselves: For retrograde orbits the
wave emission comes in a burst, while for prograde emission the
emission is a smoothly modulated oscillation, and these
characteristics increase with increasing values of $a/M$.

We emphasize that the observations above are phenomenological and
hence our explanation in Paper I of the kick/antikick cancellation is
a phenomenological one,  one
that is clearly compelling, but that does not really {\it explain} the
cancellation, since it does not explain why the prograde orbits have
slowly changing oscillations and the retrograde orbits have rapidly
changing oscillations. We offer such an explanation in this paper, and
hence show the underlying physical explanation of the linear momentum
cancellation phenomenon.

\section{Inspiral orbits}\label{sec:orbits}

The core of our explanation lies in the fact that a rotating hole
drags spacetime along with it. In the Schwarzschild spacetime, the
angular velocity $d\phi/dt$ of a particle of mass $\mu$ is
proportional to $L$, the particle's specific angular momentum
($p_\phi/\mu$), a constant of the motion.  In the Kerr spacetime,
however,
\begin{equation}\label{eq:dphidt}
\frac{d\phi}{dt}=\frac{L(1-2M/r)+2EMa/r}{E(r^2+a^2+2Ma^2/r)-2LMa/r}\,,  
\end{equation}
where $E$ is the particle's specific energy (-$p_0/\mu$), another
constant of the motion.  Due to the terms linear in $a$ in this
expression, a particle with no angular momentum can be rotating, i.e.,
can have nonzero $d\phi/dt$. It is of particular interest that for a
particle with a nonzero $L$ that has sign opposite to that of $a$, the
numerator of Eq.~(\ref{eq:dphidt}) can vanish and, since the two
cancelling terms have different $r$ dependences, can change sign as
the particle moves inward. In short, the angular velocity can reverse direction.

\begin{figure}[h]  
  \begin{center}
  \includegraphics[width=.7\textwidth ]{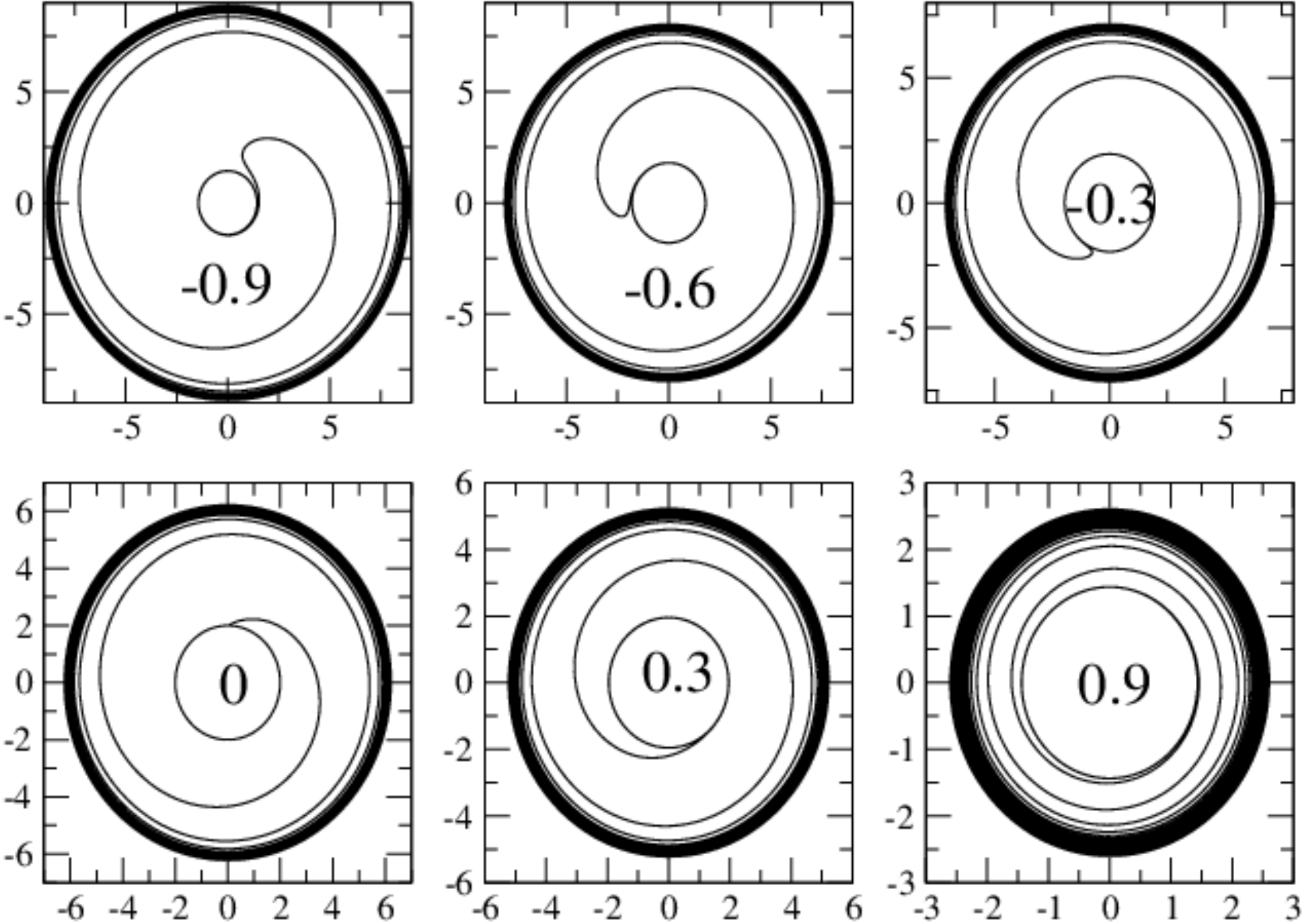}
  \caption{Plots of the equatorial particle orbits in the Kerr geometry. Each 
panel is marked with the spin parameter $a/M$. Negative numbers indicate a 
retrograde orbit; positive numbers a prograde orbit. The plots treat the $r$ and $\phi$ Boyer-Lindquist coordinates as
if they were polar coordinates in flat 2-dimensional space.
The axes show the $x=r\cos\phi$ and $y=r\sin\phi$ Cartesian-like 
coordinates based on the Boyer-Lindquist coordinates, and given in 
units of the BH mass $M$.
}
  \label{fig:all6}
  \end{center}
  \end{figure}

This reversal is clear in Fig.~\ref{fig:all6}.  The figure presents
the equatorial orbits for particles in Kerr spacetimes with various
values of the spin parameter $a/M$. Positive numbers are for prograde
infall (same sign for $L$ and $a$), and negative numbers are for retrograde orbits (opposite signs for $L$ and $a$). The plots
treat the Boyer-Lindquist\cite{BL} $r$ and $\phi$ coordinates of Kerr
spacetime as if they were 2-dimensional polar coordinates in flat
spacetime.  The dark outer band in each case indicates the particle
orbiting many times near the radius $R_{\rm ISCO}$ of the innermost
stable orbit (ISCO).  The empty circle at the center of each panel
indicates the radial coordinate location $r_{\rm hor}$ of the horizon.
It should be noticed that both the ISCO and the horizon have
different coordinate radii in different panels since these radii
depend on BH spin for a hole given mass $M$ (note the scales in use in
different panels).  The ISCO radius is quite different for prograde
and for retrograde orbits.

The trajectories are not particle geodesics. The results here use the
same radiation reaction modelling as in Ref.~\cite{SundararajanEtAlI}.  For
a particle of mass $\mu$ moving in the spacetimes of mass $M$, the loss
of orbital energy and angular momentum are second order in $\mu/M$.  It
is assumed that these losses are slow enough that the orbits can be
described as geodesics in which the particle energy and angular
momentum decrease slowly in accord with radiative losses.  For all models
reported in the current paper, the mass ratio is $\mu/M=10^{-4}$, so that
the assumption of slow rate of change is justified.

Due to this radiation reaction, the particle gradually spirals inward
from the ISCO. Once it has been driven well off the ISCO, radiation
reaction is unimportant.  Following a short transition from the
earlier adiabatic inspiral\cite{OT,BD}, the motion is negligibly different from an
infalling geodesic as the particle moves inward on a spiral that ends
at the horizon.  This motion is associated with the GW emission at $t
\to \infty$.

It is striking in Fig.~\ref{fig:all6} that the particle moving in the
prograde direction into the $a/M=0.9$ hole orbits many times and
slowly moves inward. This characteristic is less dramatic in the
$a/M=0.3$ case. 
The tendency is  yet less in the $a/M=0$ case, the orbit for a Schwarzschild
hole.
For the retrograde orbits, quite the opposite
applies; as the BH spin magnitude increases the orbit becomes less and less
dominated by circumferential motion and more and more by radial motion.

\begin{figure}[h] 
  \begin{center}
  \includegraphics[width=.4\textwidth ]{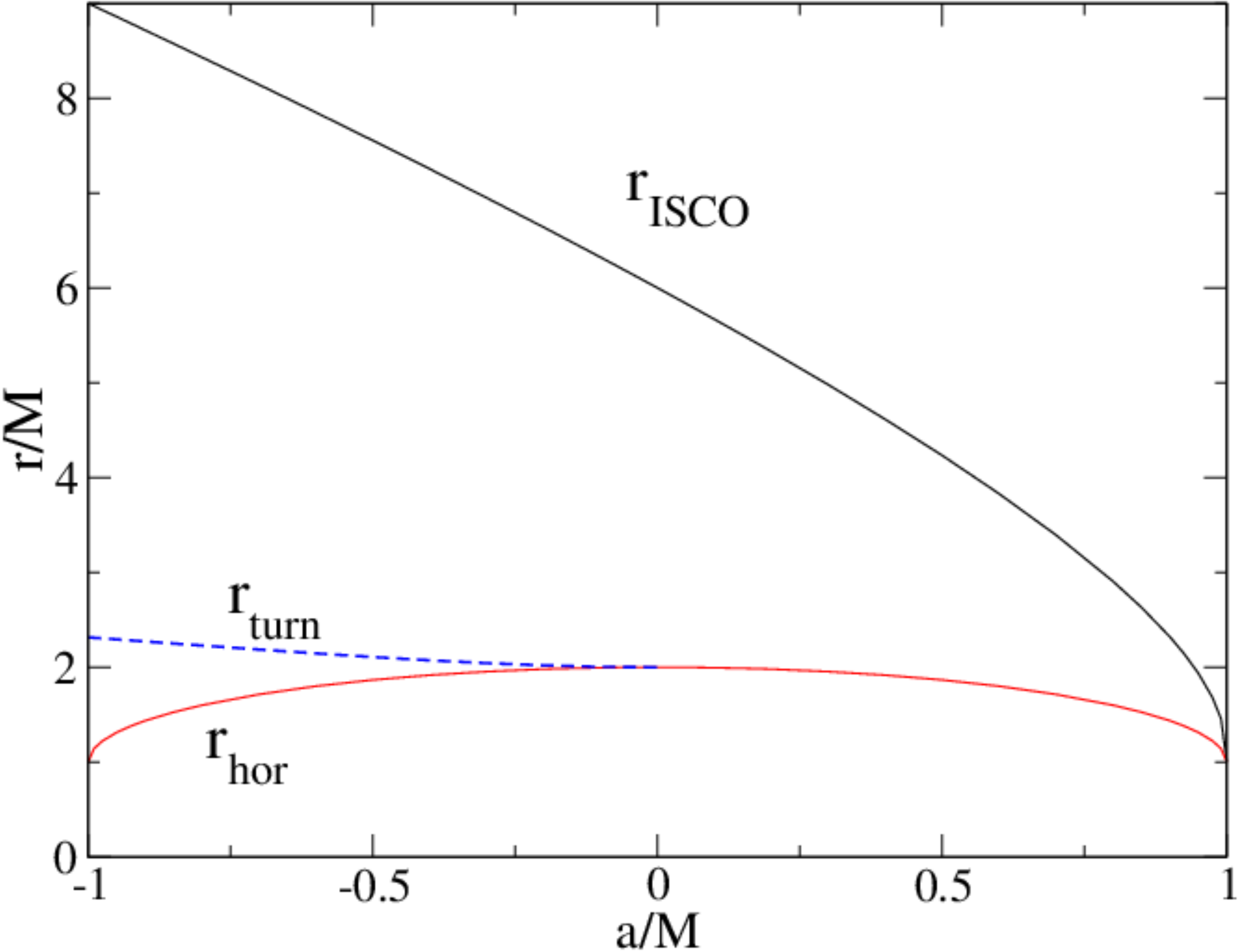}
  \caption{The radial locations of the ISCO, horizon, and turning point
are plotted as functions of the dimensionless spin parameter $a/M$. 
The turning point exists only for negative values of $a/M$, i.e., for retrograde
orbits.
}
  \label{fig:threeRs}
  \end{center}
  \end{figure}

A simple quantitative exploration of this correlation is
possible. Since radiation reaction at and interior to the ISCO is much
smaller than the secular gravitational forces (i.e., since the
trajectories are negligibly different from geodesic orbits) it is a
good approximation to set the $L$ and $E$ parameters for infall to be
those at the ISCO. These are known to give a ratio\cite{BPT}
\begin{equation}\label{eq:LbyEeq}
\frac{L}{E}=\pm\,\frac{
M^{1/2}\left(
r^2\mp2aM^{1/2}r^{1/2}+a^2
\right)
}{
r^{3/2}-2Mr^{1/2}\mp aM^{1/2}
}\,.  
\end{equation}
where the upper sign refers to prograde orbits and the lower to
retrograde.  With this ratio put into Eq. (1) we can find, for
retrograde orbits, the approximate radial location $r_{\rm turn}$ at
which $d\phi/dt$ changes sign during the plunge.
These locations
are presented, as functions of $a/M$ in Fig.~\ref{fig:threeRs} along with radial locations
of the ISCO and horizon.
\vspace{6pt}

\begin{figure}[h]
  \begin{center}
  \includegraphics[width=.5\textwidth ]{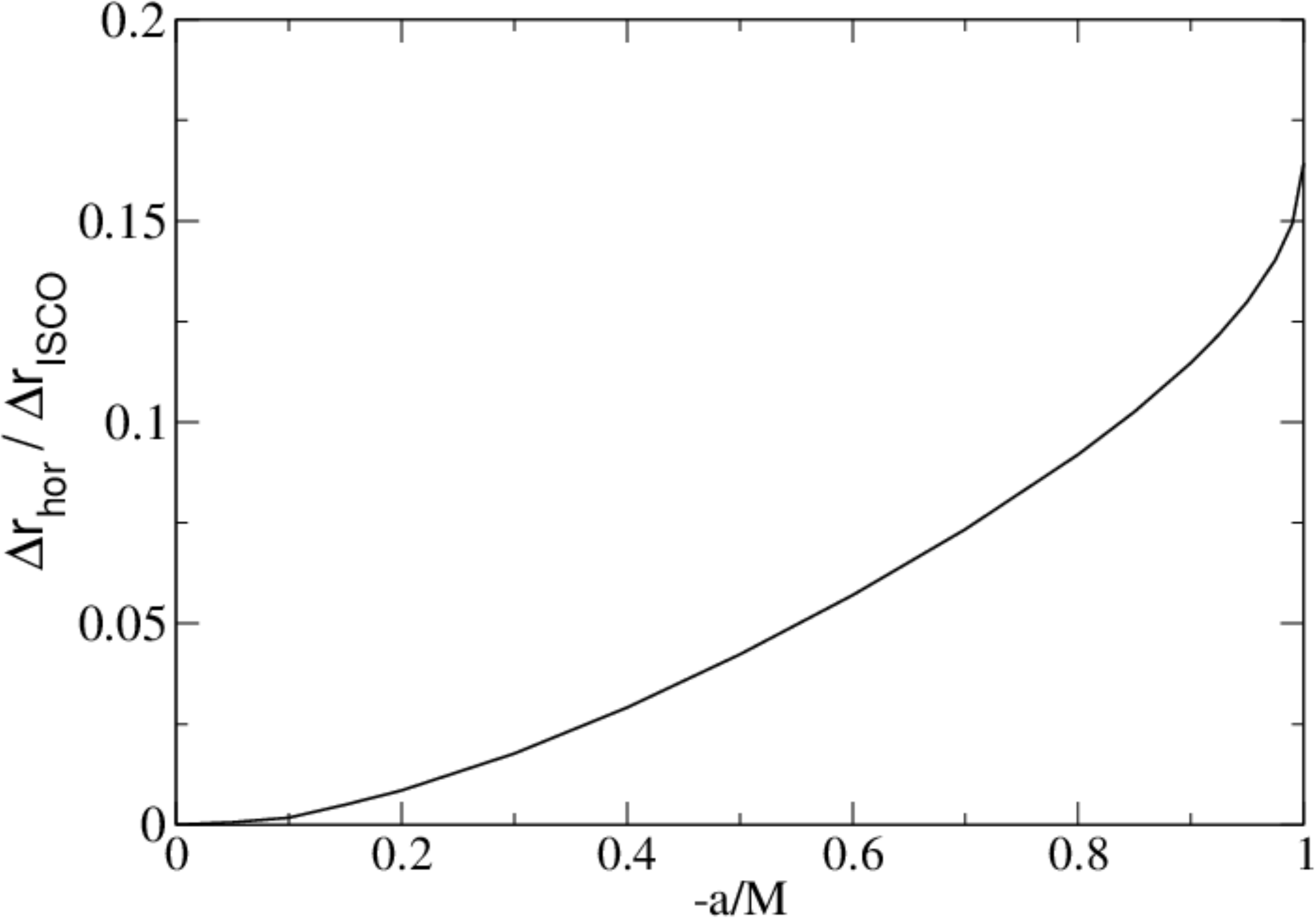}
  \caption{For retrograde inspiral, the fraction of ISCO to horizon radius for which the angular velocity is
reversed. Here $\Delta r_{\rm turn}$ is $r_{\rm turn}-r_{\rm hor}$, the radial distance
from the turning point to the horizon, and $\Delta r_{\rm ISCO}$ is $r_{\rm ISCO}-r_{\rm hor}$, the radial distance from ISCO to horizon.
}
  \label{fig:DeltaRatio}
  \end{center}
  \end{figure}

If the angular velocity reversal occurs too close (in some sense) to
the horizon, gravitational redshift effects dominate to suppress
outgoing gravitational wave energy and momentum. A crude index of the
importance of the angular velocity reversal is therefore the ratio of
the reversed-motion radial span $r_{\rm turn}-r_{\rm hor}$ to the full
radial span $r_{\rm ISCO}-r_{\rm hor}$. This ratio is shown, for
retrograde orbits, in Fig.~\ref{fig:DeltaRatio}.
The implications of Fig.~\ref{fig:all6} are supported by the 
results in this figure; the importance of the angular velocity reversal for
retrograde infall increases dramatically with increasing BH spin.

We have so far focused on the retrograde orbits, while it had been the
high spin
prograde orbits, that produced the most interesting cancellation phenomenon.
 We now understand this to be due to the gradual
orbiting for prograde cases after the particle has detached from the ISCO
and is spiralling in toward the horizon. This gradual spiralling is
particularly clear for the prograde inspiral with $a/M=0.9$ shown in
Fig.~\ref{fig:all6}.  A suggestion of the physical basis for this
can be seen in Eq.~(\ref{eq:dphidt}): for prograde orbits, in which $L$
and $a$ have the same sign, the two terms in the numerator of
$d\phi/dt$ have the same sign, while for retrograde orbits they would
have opposite signs. This suggests that $d\phi/dt$ is larger in the
prograde case and that it increases with increasing BH spin.

\begin{figure}[h]
  \begin{center}
  \includegraphics[width=.5\textwidth ]{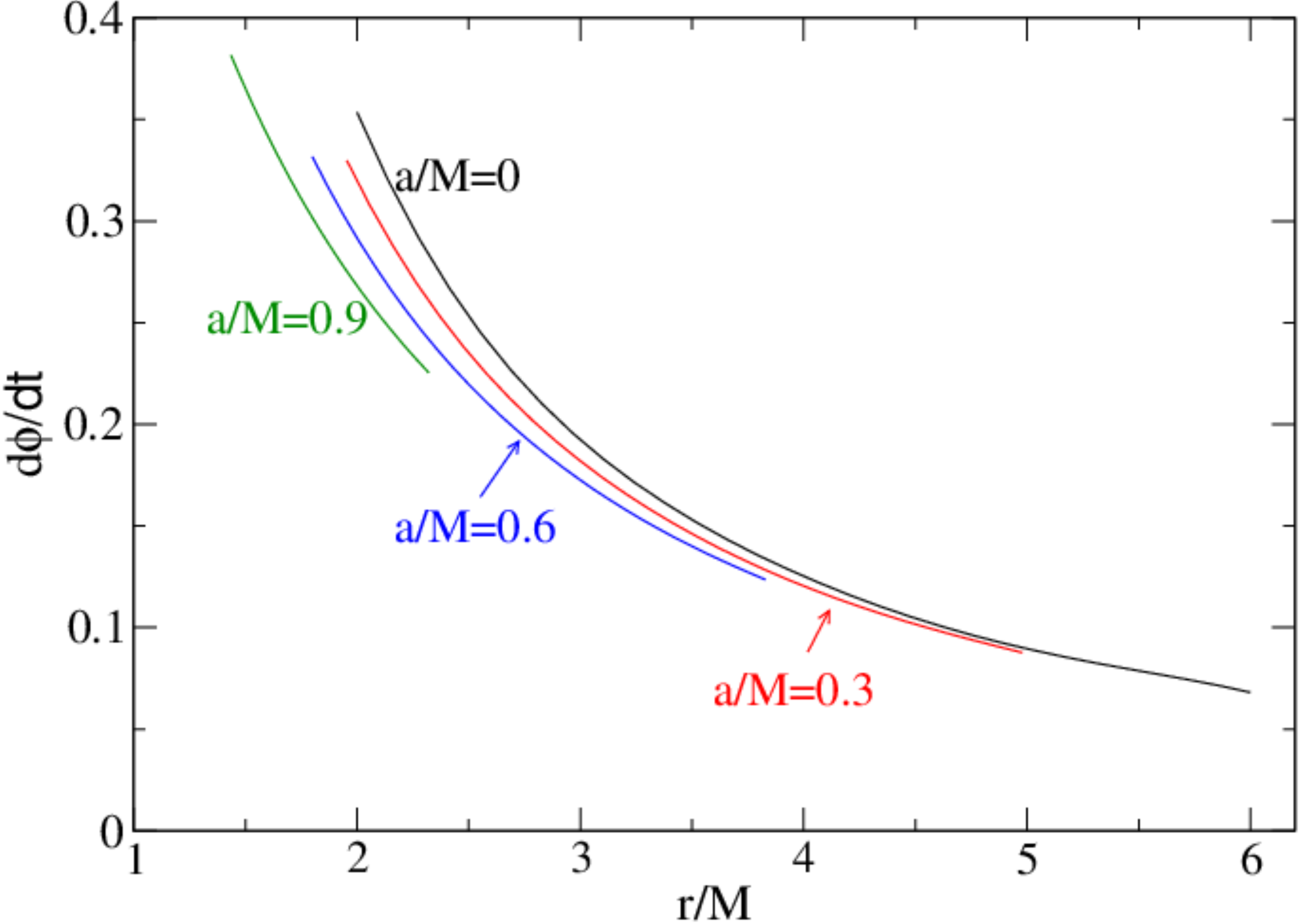}
  \caption{The values of the particle angular velocity $d\phi/dt$
as it spirals from the ISCO to the horizon.
}
  \label{fig:progdphidt}
  \end{center}
  \end{figure}

The situation is actually rather more complicated. For one thing,
$d\phi/dt$ for the inspiral depends on radius; the particle whirls
faster (as measured in coordinate time) as it approaches the
horizon. This is shown
 in Fig.~\ref{fig:progdphidt}, along with the dependence of
angular velocity on $a/M$. We should not
lose sight of the fact that $d\phi/dt$ by itself does not really
determine the kick/antikick cancellation.  Rather, the important point
is the way in which the amplitude of linear momentum flux changes
slowly for particle motion after the plunge, i.e., inside the ISCO.
Figure \ref{fig:progdphidt} is therefore only mildly suggestive of the
reason for the increase in cancellation with increasing $a/M$.
\vspace{6pt}

\section{Tests of the orbit-dominance explanation}\label{sec:tests}
\subsection{Kerr orbits embedded in Schwarzschild spacetime}\label{subsec:KinS}

Here we test the hypothesis that the nature of the kick, and of
gravitational wave emission more generally, is dominated by the nature
of the trajectory, rather than by the nature of the spacetime in which
the gravitational waves are generated and through which they
propagate. One way of investigating what dominates, trajectory or
spacetime, is to take a trajectory from, say, spacetime A, put it as a
source in spacetime B, and see whether the emerging radiation is
characteristic of the trajectory or the spacetime.  This procedure
amounts to putting into spacetime B an orbit that differs dramatically
from a geodesic orbit in spacetime A. This configuration, then, cannot
be considered to be the extreme mass ratio limit of a process in
general relativity. Nevertheless, it is mathematically consistent in
linear particle perturbation theory, since the specification of the
source in such calculations is an independent step.

\begin{figure}[h]
  \begin{center}
  \includegraphics[width=.42\textwidth ]{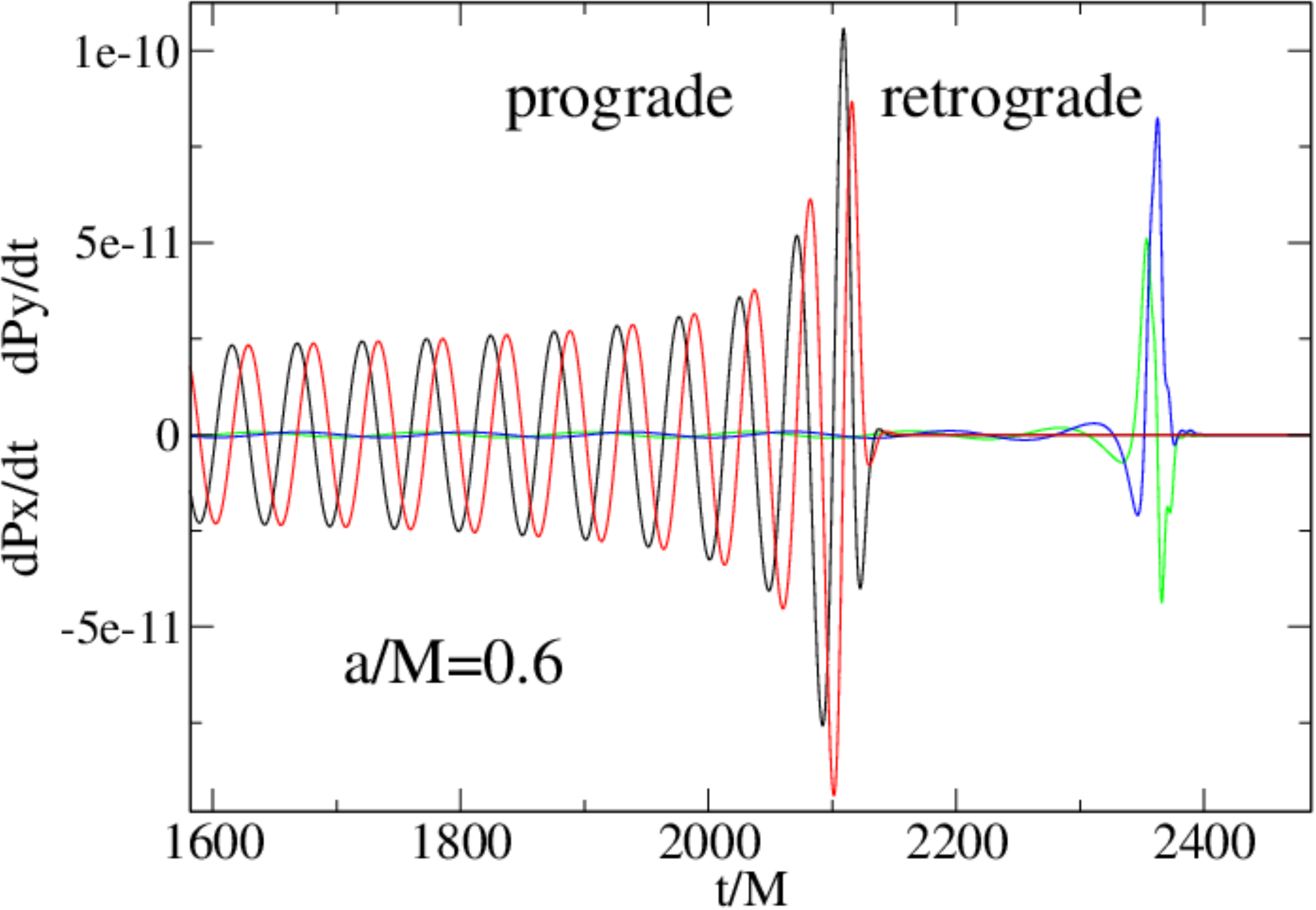}\hspace{.02\textwidth}
  \includegraphics[width=.4\textwidth ]{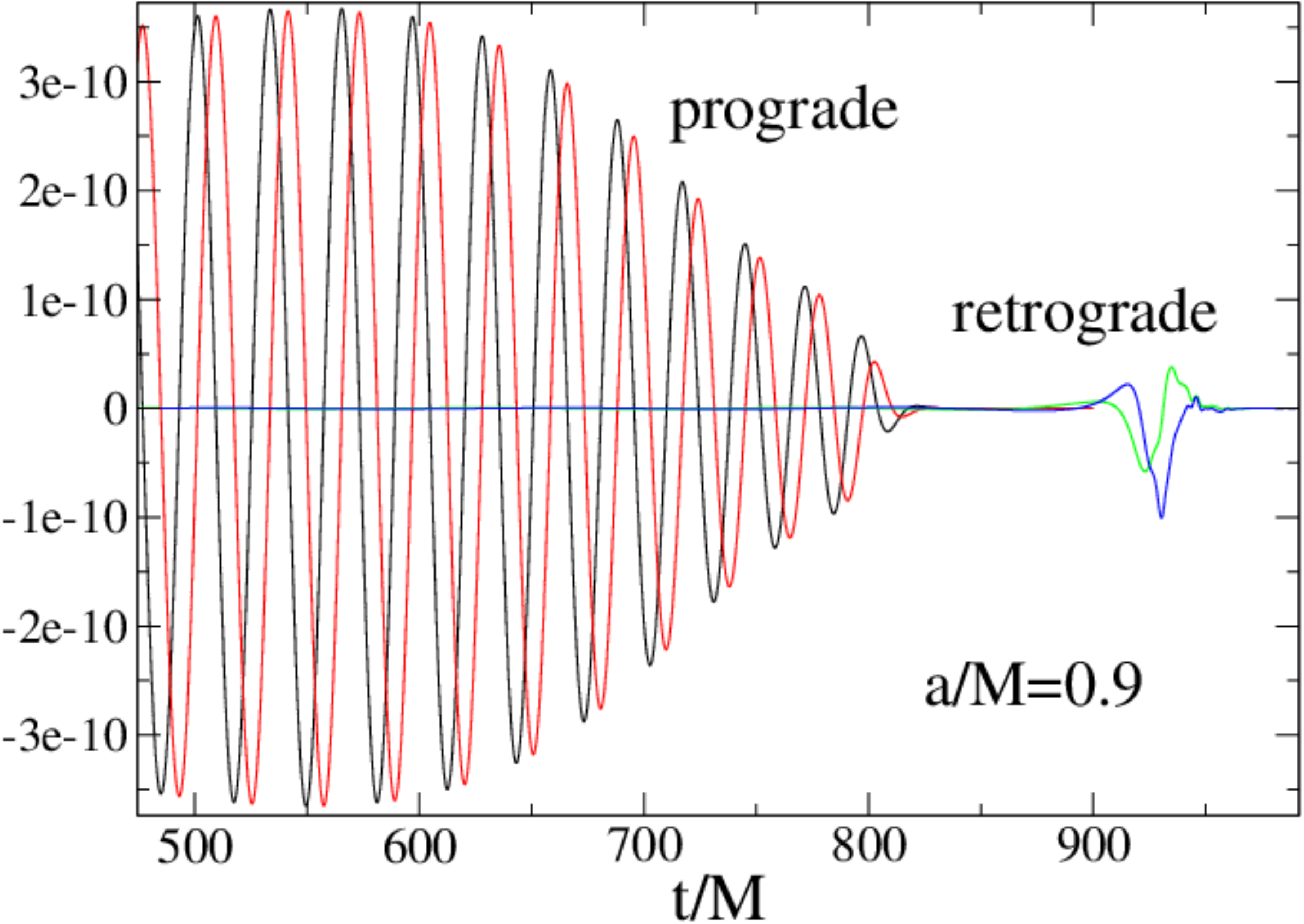}
  \caption{
The two components of momentum flux generated by Kerr inspiral orbits 
 placed in a Schwarzschild background. The plots on the left show trajectories
for $a/M=0.6$; those on the right correspond to  
$a/M=0.9$. (Note the different scales on the two plots; 
the amplitude of the retrograde burst for
$a/M=0.9$ is slightly
higher than than that for  $a/M=0.6$.)}
  \label{fig:Schwkickcomparo}
  \end{center}
  \end{figure}

\begin{figure}[h]
  \begin{center}
  \includegraphics[width=.4\textwidth ]{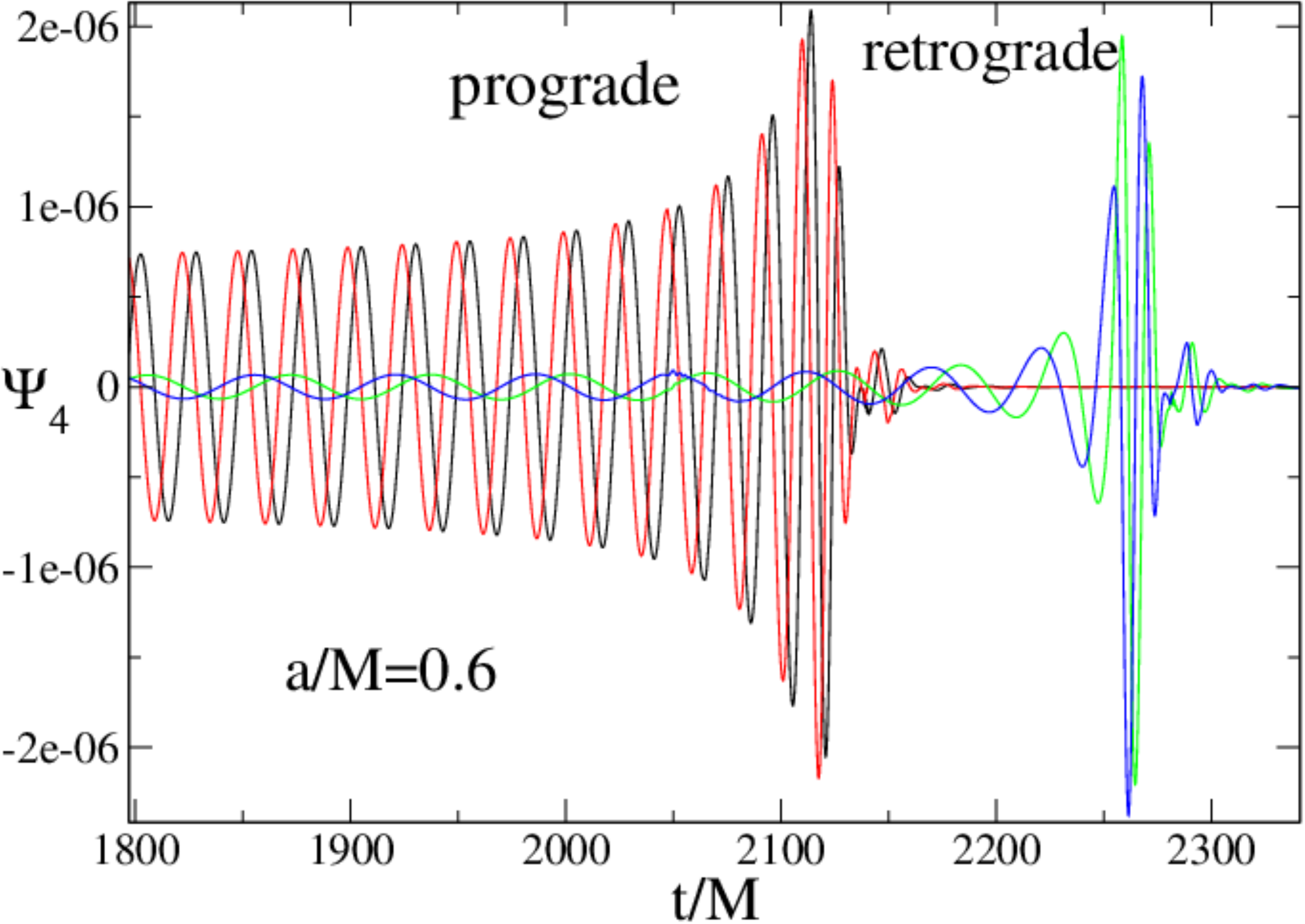}\hspace{.02\textwidth}
  \includegraphics[width=.4\textwidth ]{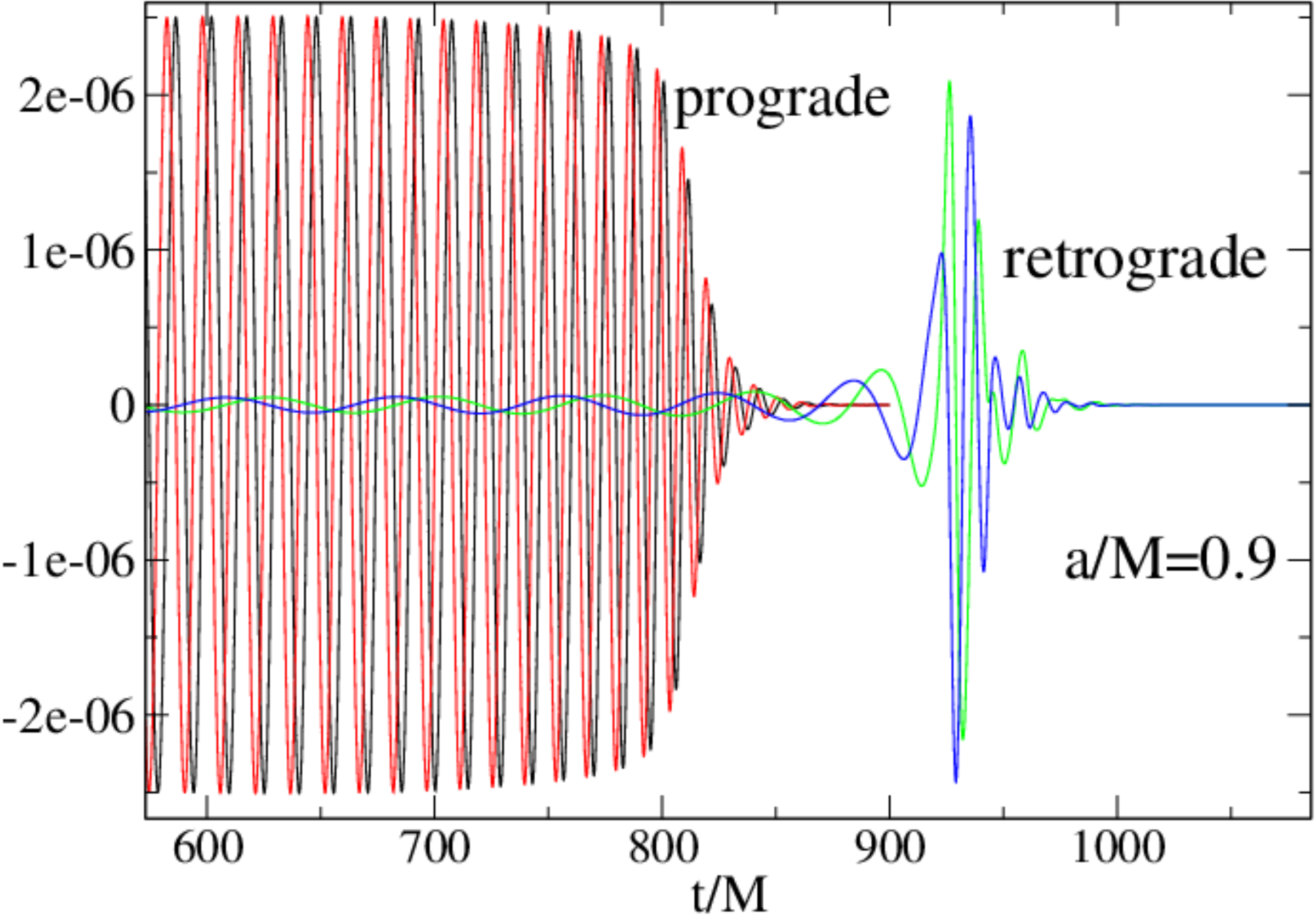}
  \caption{The real and imaginary parts of the  $m=2$ component of the  Teukolsky 
wave function $\Psi_4$ from   Kerr inspiral orbits 
 placed in a Schwarzschild background. The plots on the left show trajectories
for $a/M=0.6$; those on the right correspond to  
$a/M=0.9$.}
  \label{fig:Schwavecomparo}
  \end{center}
  \end{figure}

The results of tests of this type are shown in
Figs.~\ref{fig:Schwkickcomparo} and \ref{fig:Schwavecomparo}. The
plots show the components 
$dP_x/dt$ and $dP_y/dt$
of gravitational wave momentum flux
 from equatorial Kerr orbits placed in a
Schwarzschild spacetime.  In principle, one can start with the Kerr
trajectory for a hole of mass $M$ and nonzero spin parameter $a/M$. One
then uses the coordinate functions $r(t)$ and $\phi(t)$ in
Boyer-Lindquist coordinates as the specification of an orbit in the
Schwarzschild geometry of the same mass $M$.  In practice, this procedure
enounters a problem at the horizon, since the radial location of the
Kerr horizon $r=M+\sqrt{M^2-a^2\;}$ is less than the radial position
$2M$ for the Schwarzschild geometry. The Kerr trajectory coordinate
specification would therefore extend inside the horizon in the
Schwarzschild geometry.

This problem is avoided by matching not the radii of the two inspiral
trajectories, but rather their values of the function $\Delta_a = r^2
- 2Mr + a^2$.  Given a trajectory $[r_K(t), \phi(t)]$ in the Kerr
geometry, we map this to a trajectory $[r_S(t), \phi(t)]$ by requiring
that $\Delta_a(r_K) = \Delta_0(r_S)$ at each moment $t$.  
In this way the horizon location (at $\Delta=0$)
of one spacetime corresponds to the horizon (at $\Delta=0$) in the other.

\vspace{6pt}

\vspace{6pt}

The resulting plots, shown in Figs.~\ref{fig:Schwkickcomparo} and
\ref{fig:Schwavecomparo}, strongly support the notion of trajectory
dominance.  Figure~\ref{fig:Schwkickcomparo} shows the momentum flux
components from $a/M=0.6$ and $a/M=0.9$ orbits embedded in the
Schwarzschild spacetime. For both spins there is a dramatic difference
between the prograde and the retrograde momentum fluxes. A comparison
of Fig.~\ref{fig:Schwkickcomparo} with Figs.~\ref{fig:pandpdot06} and
\ref{fig:pandpdot09} shows, moreover, that the qualitative nature of
the momentum emission from any of the Kerr orbits in Schwarzschild is
the same as the that in the Kerr spacetime in which the orbits are
approximate geodesics.  Figure \ref{fig:Schwavecomparo} makes the same
comparison for the gravitational waves, in particular for the $m=2$ component
of the Teukolsky function $\Psi_4$. Again, the gravitational wave
emission for prograde orbits are dramatically different from
retrograde orbits, and the emission is qualitatively the same for a
Kerr orbit in the Schwarzschild spacetime as it is in the spacetime in 
which it is an approximate geodesic.

  \begin{figure}[h]
  \begin{center}
  \includegraphics[width=.45\textwidth ]{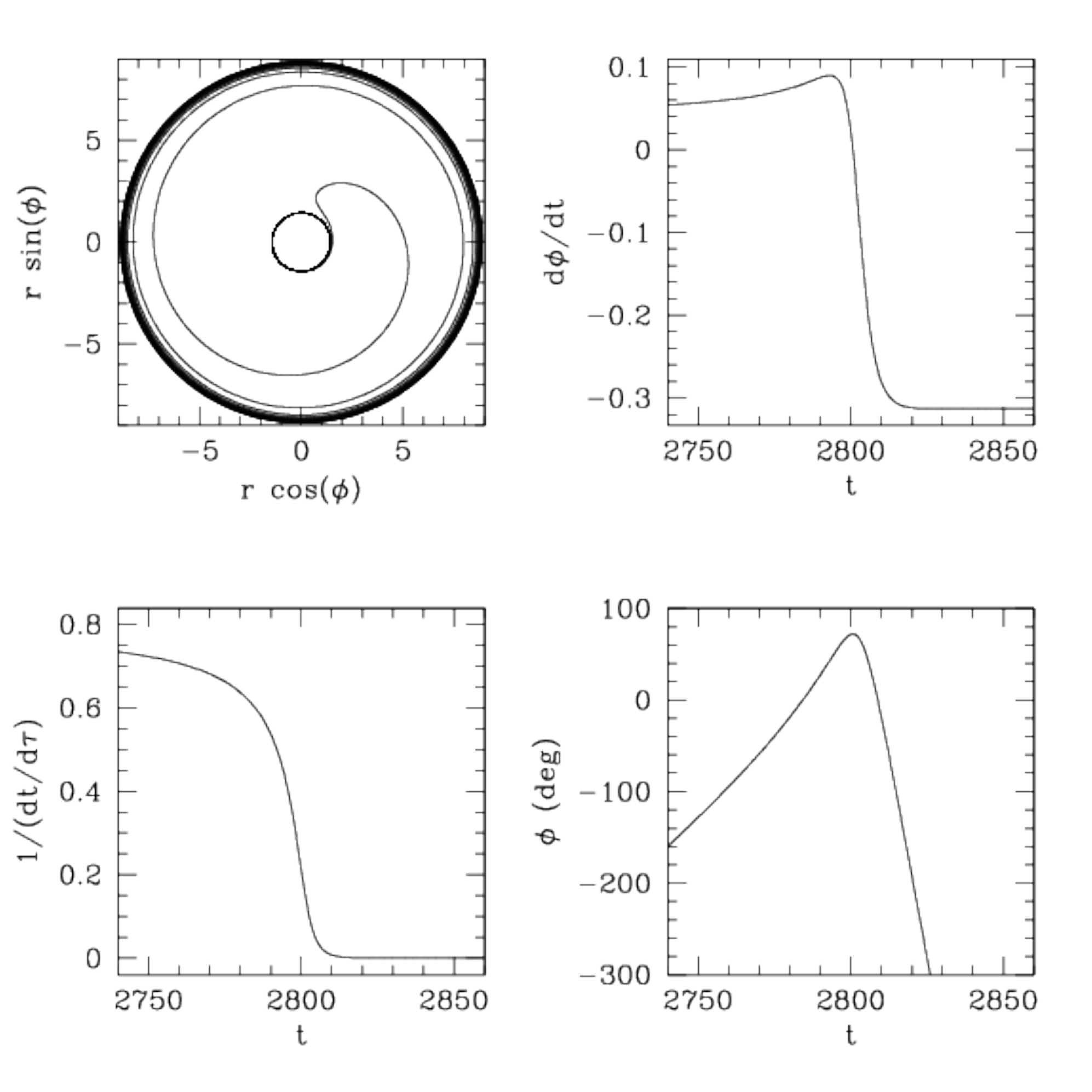}\hspace{.01\textwidth}
\includegraphics[width=.45\textwidth ]{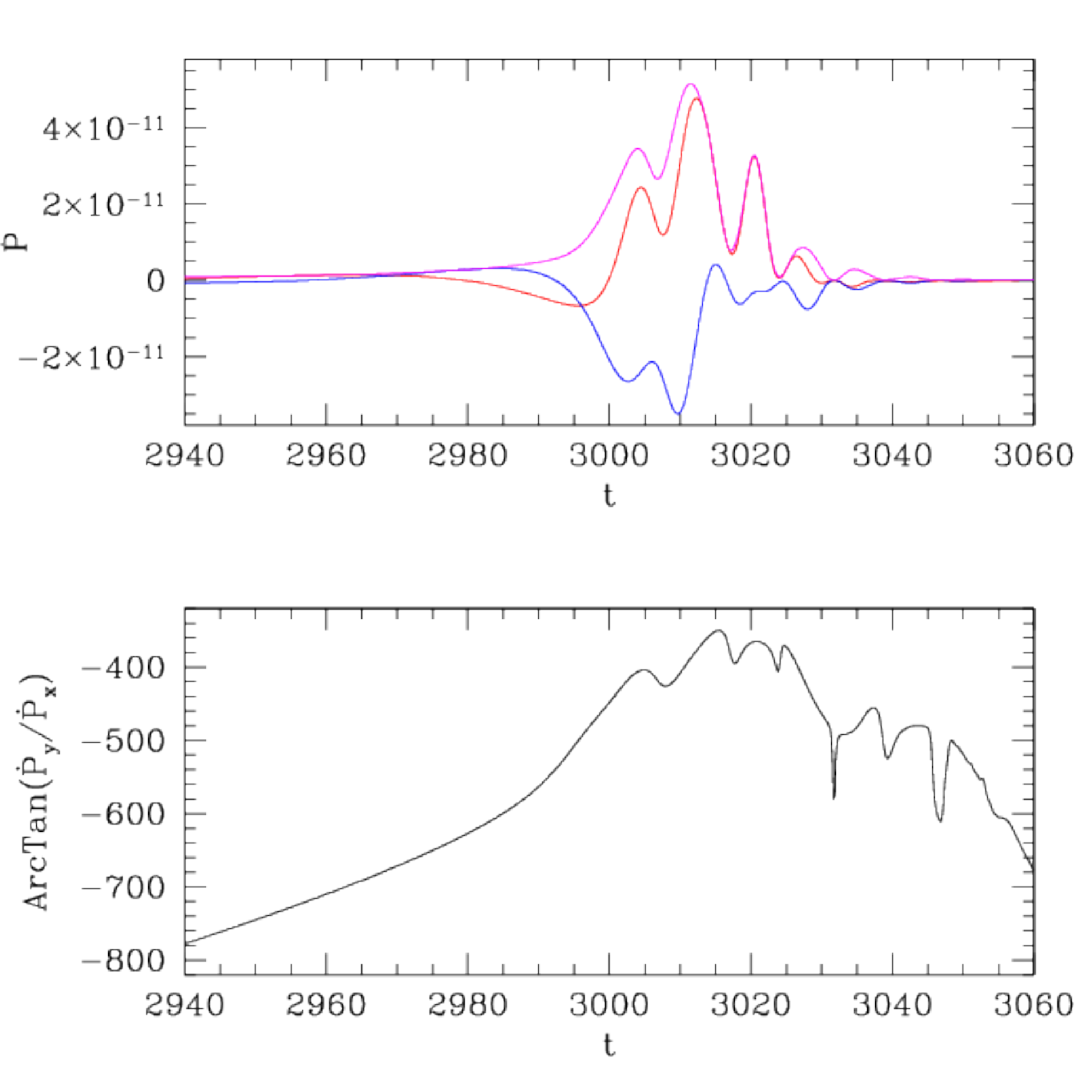}
  \caption{Trajectory and flux for the retrograde inspiral into a Kerr BH
with $a/M=0.9$. See text for details.}
  \label{fig:trajnflux09}
  \end{center}
  \end{figure}

  \begin{figure}[h]
  \begin{center}
  \includegraphics[width=.45\textwidth ]{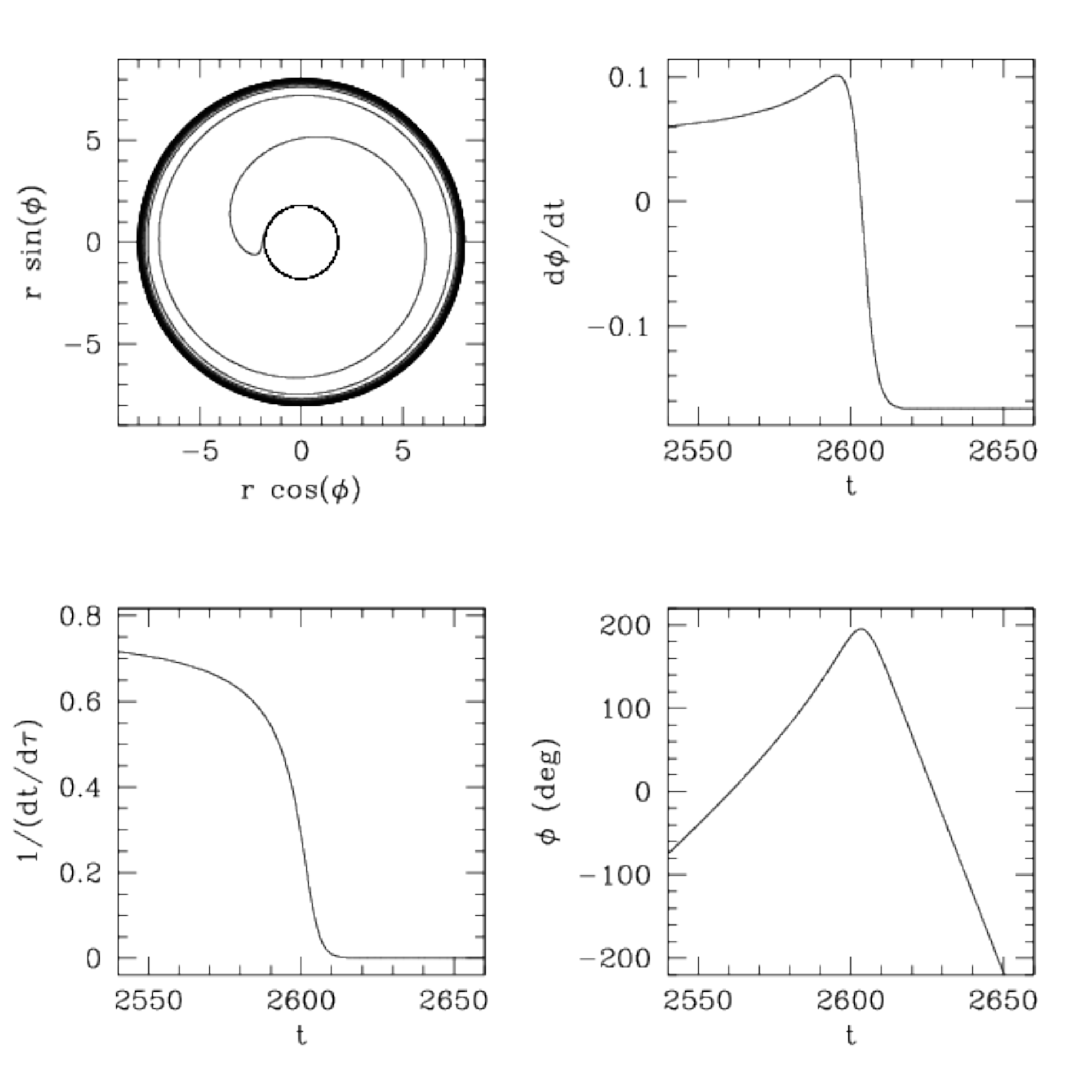}\hspace{.01\textwidth}
\includegraphics[width=.45\textwidth ]{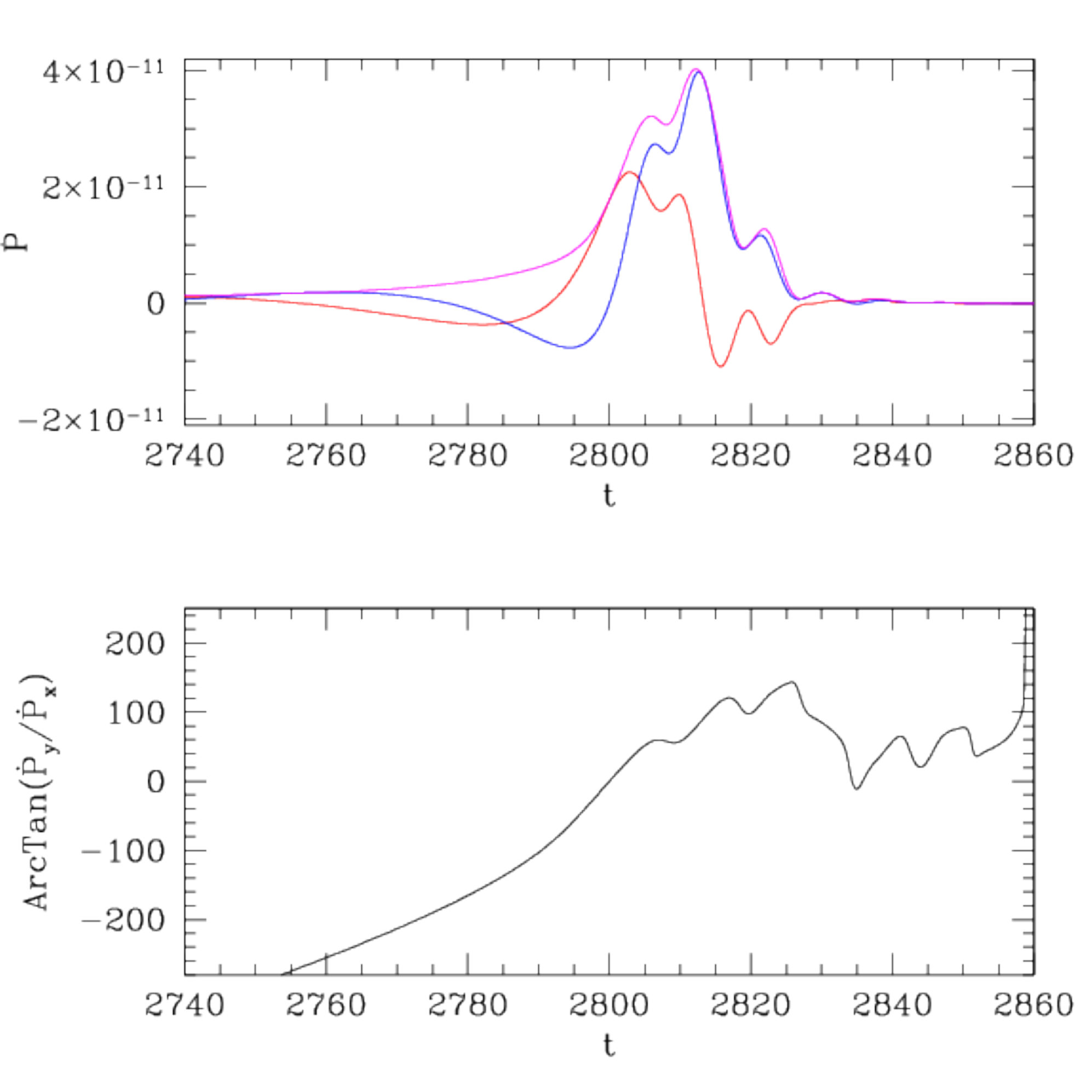}
  \caption{Trajectory and flux for the retrograde inspiral into a Kerr BH
with $a/M=0.6$. See text for details.}
  \label{fig:trajnflux06}
  \end{center}
  \end{figure}

  \begin{figure}[h]
  \begin{center}
  \includegraphics[width=.45\textwidth ]{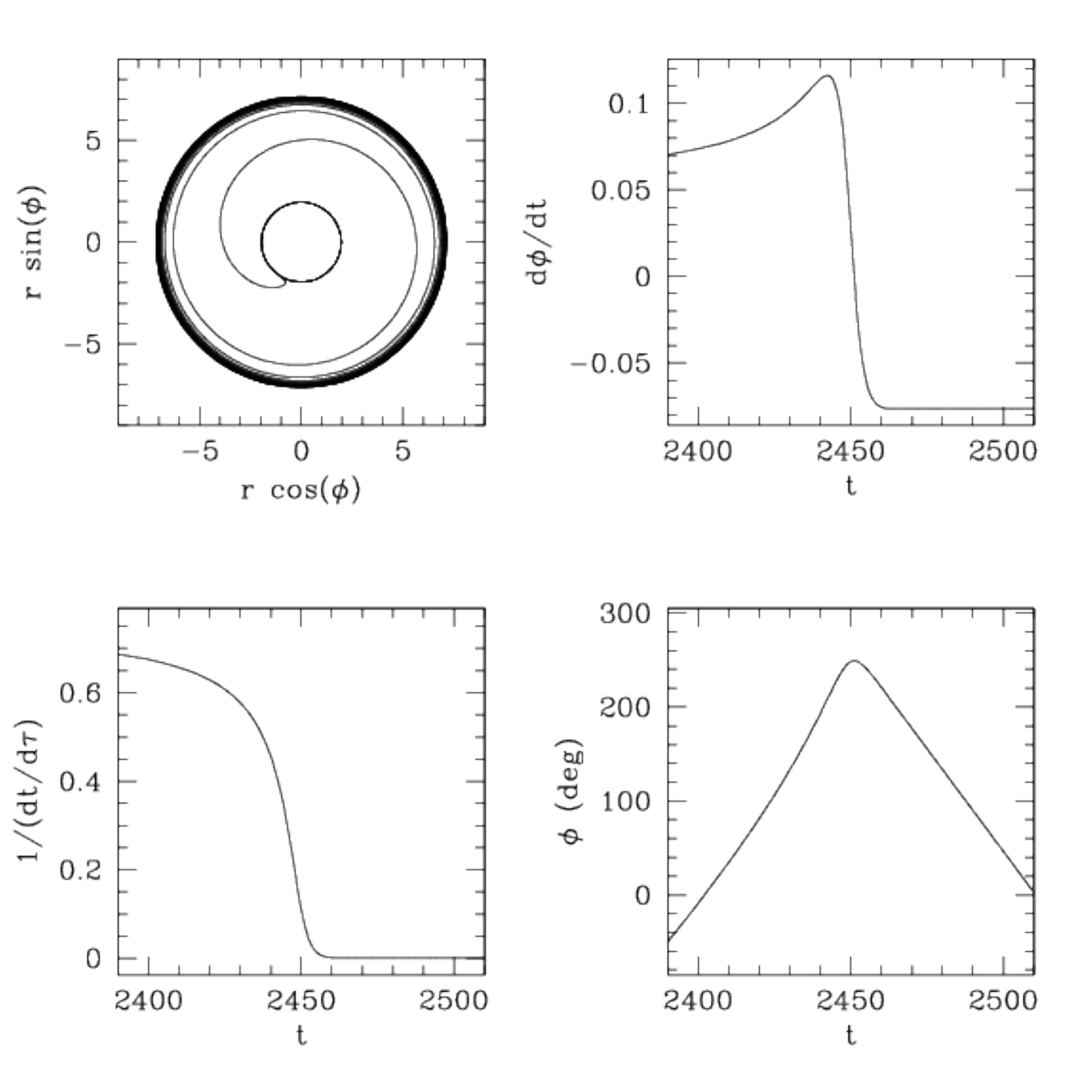}\hspace{.01\textwidth}
\includegraphics[width=.45\textwidth ]{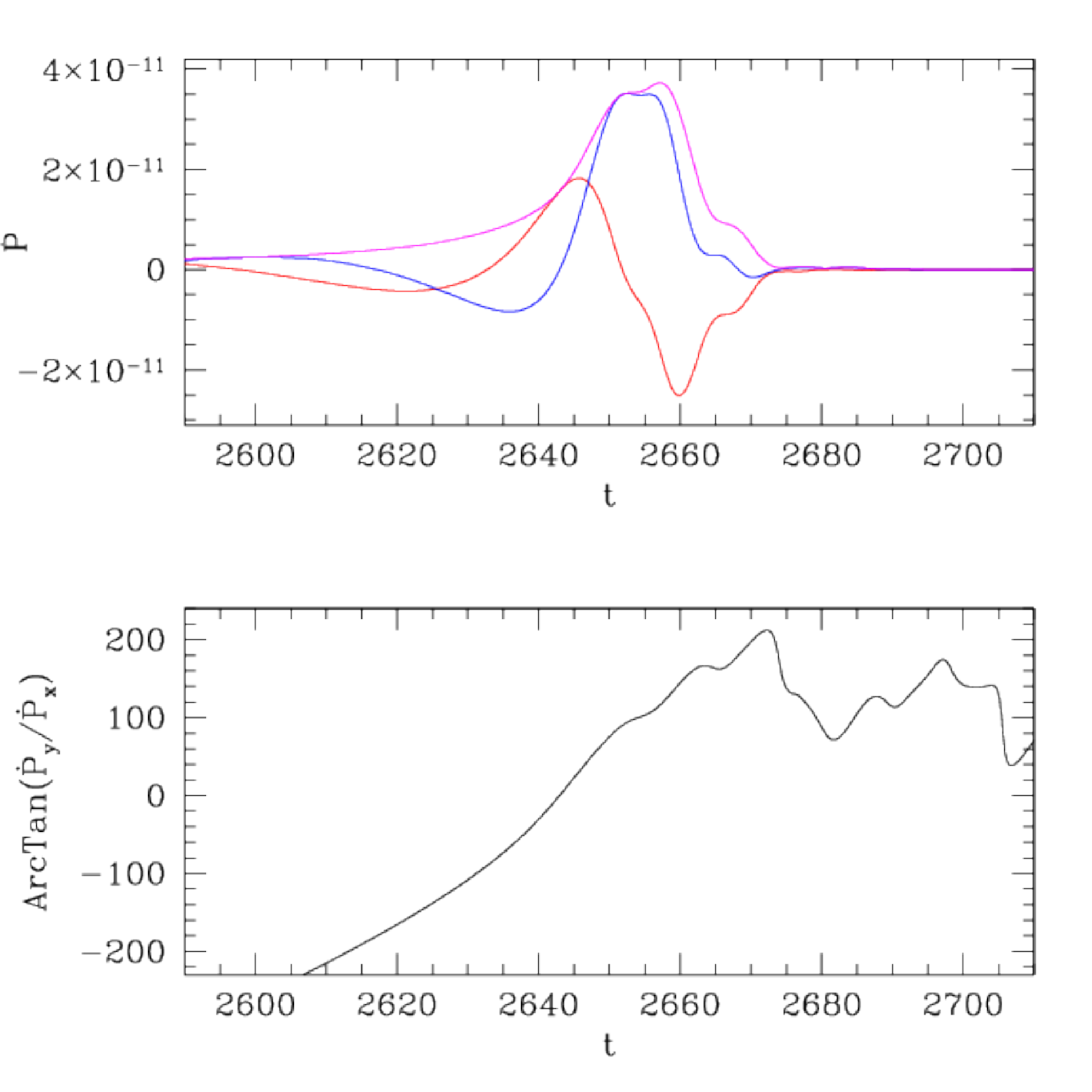}
  \caption{Trajectory and flux for the retrograde inspiral into a Kerr BH
with $a/M=0.3$. See text for details.}
  \label{fig:trajnflux03}
  \end{center}
  \end{figure}

\subsection{Correlations of timing and angular direction}\label{subsec:timing}

As a very different test of trajectory dominance we look for features
of the emerging radiation that can be correlated with features of the
orbits. In particular, a strong argument for trajectory dominance was
made in Sec.~\ref{sec:orbits} based on the post-plunge reversal of
angular velocity for retrograde orbits. Here we look at evidence that the burst
of gravitational radiation, and especially of linear momentum, from retrograde
orbits really does come from that reversal event.

The results, shown Figs.~\ref{fig:trajnflux09}, \ref{fig:trajnflux06},
and \ref{fig:trajnflux03} compare features of the
orbit, in the left two panels, with results, in the right panels, for
momentum flux observed at a Boyer-Lindquist radius of $200\,M$. In
Fig.~\ref{fig:trajnflux09}, for the retrograde inspiral orbit into a
Kerr BH with $a/M=0.9$, the left panel repeats the corresponding
panel in Fig.~\ref{fig:all6}, showing a picture of the trajectory and
showing the angular velocity reversal occurring fairly close to the
horizon. The plots of azimuthal angle $\phi$ and angular velocity
$d\phi/dt$ confirm that reversal takes place around $\phi=80^\circ$,
and indicate that this occurs at coordinate time
$t\approx2800\,M$. The plot of $1/(dt/d\tau)=1/\gamma$ shows
the relationship of  coordinate time and particle proper time, and hence
shows the development of the redshift factor $\gamma$. The story told
by these plots then is that the relativistic effects increase after the 
plunge, are
fairly strong around the time of angular reversal, and show the
subsequent redshifting away as the particle asymptotically disappears
in the horizon.

The right panel shows features of the linear momentum flux from this orbit. 
The plot of momentum fluxes (as in 
Fig.~\ref{fig:pandpdot09}) show that the burst of linear momentum, starting
around $t=3000\,M$ is in reasonably good time agreement with the time of 
orbital angular velocity reversal at $t=2800\,M$, when allowance is made for
propagation time to the observation radius $r=200\,M$.

The right panel also presents $\arctan(\dot P_y/\dot P_x)$, which
gives an estimate of the direction in which the linear momentum is
radiated.  The direction of the momentum differs substantially from
the angle at which the reversal of the angular velocity takes place.
This is not too surprising, bearing in mind that due to frame
dragging and strong-field propagation effects the radiation does not
proceed outward in a constant $\phi$ direction.  What is telling is
the time development of the angle of the radiated momentum.  That
angle increases monotonically to $t \approx 3000 M$, which mirrors
(with an appropriate shift) the time at which the reversal occurs.
Beyond this, the angle associated with the flux is roughly constant or
even decreasing (bearing in mind that the flux decays rapidly, and
this angle is likely to be dominated by numerical noise as we go
forward in time).

Figures \ref{fig:trajnflux06} and \ref{fig:trajnflux03} show analogous
results for the retrograde inspiral into holes of $a/M=0.6$ and
$0.3$. The discussion of the $a/M=0.9$ case applies to these as well. The differences
are only in a few of the details.

\section{Conclusion}\label{sec:conc}

For particles in BH spacetimes, the spacetime itself has two effects:
it determines the trajectories of the particles, and it governs the
radiation emerging from the motion of those particles. We have given
evidence, evidence that we consider compelling, that it is the trajectories
that are crucial to the nature of the  emerging radiation.  The case
was built in Secs.~\ref{sec:orbits} and \ref{sec:tests}.
In those sections it was argued that aside from determining
the particle trajectories,  the main role of the
  structure of the spacetime {\em per se} is to cut off the particle
  generated radiation as the particle asymptotes to the horizon. We've
  seen that especially in Figs.~\ref{fig:trajnflux09},
  \ref{fig:trajnflux06}, \ref{fig:trajnflux03}.

Although the work reported in this paper was originally motivated by a
phenomenon of prograde orbits, the kick/antikick cancellation, it
turns out that it is the retrograde orbits that are the more
interesting, and provide the strongest evidence for the dominant role
of the the particle orbits. This is due, in particular, to the angular
velocity reversal, a feature of the orbit that can be directly
connected to the pattern of radiation.

An interesting working hypothesis is that in a broader class of IPM
models the radiation from a particle in a BH background can usefully
be broken down into spacetime $\rightarrow$ trajectory $\rightarrow$
radiation in which the spacetime plays a role in the last step only
through the horizon cut off.  Although such a simplifying picture is
most applicable to the EMRI limit, we note that the kick/antikick
cancellations exhibited for comparable mass holes in the work of
Schnittman {\em et al.} indicate that this picture must be at least
partially applicable for comparable mass holes.

We are now using the idea of trajectory dominance to look for a deeper
understanding of generation of radiation during the plunge, the most
important epoch of the IPM, but the epoch which is most difficult to
treat with simple approximations.


\section*{Acknowledgments}

RHP gratefully acknowledges support of this work by the UTB Center for Advanced 
Radio Astronomy
GK acknowledges research support from NSF Grant Nos. PHY-1016906, PHY-1135664
and PHY-1303724.
This work was supported at MIT by NSF Grant PHY-1068720. SAH also
gratefully acknowledges fellowship support by the John Simon
Guggenheim Memorial Foundation, and sabbatical support from the
Canadian Institute for Theoretical Astrophysics and the Perimeter
Institute for Theoretical Physics


\begin{thebibliography}{11}
\expandafter\ifx\csname natexlab\endcsname\relax\def\natexlab#1{#1}\fi
\expandafter\ifx\csname bibnamefont\endcsname\relax
  \def\bibnamefont#1{#1}\fi
\expandafter\ifx\csname bibfnamefont\endcsname\relax
  \def\bibfnamefont#1{#1}\fi
\expandafter\ifx\csname citenamefont\endcsname\relax
  \def\citenamefont#1{#1}\fi
\expandafter\ifx\csname url\endcsname\relax
  \def\url#1{\texttt{#1}}\fi
\expandafter\ifx\csname urlprefix\endcsname\relax\def\urlprefix{URL }\fi
\providecommand{\bibinfo}[2]{#2}
\providecommand{\eprint}[2][]{\url{#2}}


\bibitem{vgm10} M.\ Volonteri, K.\ G\"ultekin, 
and M.\ Dotti, Mon.\ Not.\ R.\ Astron.\ Soc.\ {\bf 404}, 2143 (2010); arXiv:1001.1743

\bibitem{zrdzp10} O.\ Zanotti, L.\ Rezzolla, L.\ Del Zanna, and C.\ Palenzuela, Astron.\ and Astrophys.\ {\bf 523}, A8 (2010); arXiv:1002.4185.

\bibitem{gmmc11} J.\ Guedes, P.\ Madau, L.\ Mayer, and S.\ Callegari, Astrophys.\ J.\ {\bf 729}, 125 (2011); arXiv: 1008.2032.

\bibitem{ssh11} D.\ Sijacki, V.\ Springel, 
and M.\ Haehnelt, Mon.\ Not.\ R.\ Astron.\ Soc., in press; arXiv:1008.3313.

\bibitem{bclh11} L.\ Blecha, T.\ J.\ Cox, A.\ Loeb, and L.\ Hernquist, Mon.\ Not.\ R.\ Astron.\ Soc., in press; arXiv:1009.4940.



\bibitem{runawayAGNs}	
S. Komossa, H. Zhou, H. Lu, Astrophys. J. {\bf678}, L81 (2008);
G. Shields {\it et al.}, Astrophys.\ J. {\bf707}, 936 (2009);
P. G. Jonker {et al.}, Mon.\ Not.\ R.\ Astron.\ Soc.\ {\bf407}, 645 (2010);
D.\ Batcheldor, A.\ Robinson, D.\ J.\ Axon, E.\ S.\ Perlman, D.\ Merritt, 
Astrophys.\ J. {\bf717}, L6 (2010);
L.~Blecha, F.~Civano, M.~Elvis, and A.~Loeb, 
Mon.\ Not.\ R.\ Astron.\ Soc.\ {\bf428}, 1341 (2013),
\eprint{arXiv:astro-ph/1205.6202}.

\bibitem{bigkicks} J.\ G. Baker, J. Centrella, D. Choi, M. Koppitz,
  J. R. van Meter, M. C. Miller, Astrophys. J. {\bf653}, L93 (2006);
  M. Campanelli, C. O. Lousto, Y. Zlochower and D. Merritt,
  Astrophys. J. {\bf659}, L5 (2007); J. A. Gonzalez, M. D. Hannam,
  U. Sperhake, B. Bruegmann and S. Husa, Phys. Rev. Lett. {\bf98}, 231101
  (2007); M. Campanelli, C. O. Lousto, Y. Zlochower and D. Merritt,
  Phys. Rev. Lett. {\bf98}, 231102 (2007); J. G. Baker, W. D. Boggs,
  J. Centrella, B. J. Kelly, S. T. McWilliams, M. C. Miller, J. R. van
  Meter, Astrophys. J {\bf682}, L29 (2008); B. Bruegmann, J. A. Gonzalez,
  M. D. Hannam, S. Husaand U. Sperhake, Phys. Rev.\ D {\bf77}, 124047 (2008);
  C. O. Lousto, Y. Zlochower, Phys. Rev. Lett. {\bf107}, 231102 (2011).



\bibitem{visualization}
R.~Owen {\em et al.}, ``Frame-Dragging Vortexes and Tidal Tendexes Attached to Colliding Black Holes: Visualizing the Curvature of Spacetime,"
{Phys.~Rev.~Lett.}, {\bf106},  {151101} ({\tt arXiv:1012.4869}) (2011);
D.~A.~{Nichols}, {\it et al.},
``{Visualizing spacetime curvature via frame-drag 
vortexes and tidal tendexes: General theory and weak-gravity 
applications}," {Phys.~Rev.~D} {\bf84}, 124014 ({\tt arXiv:1108.5486}) (2011);
A.~{Zimmerman}, D.~A.~{Nichols},  and F.~{Zhang},
``{Classifying the isolated zeros of asymptotic gravitational radiation by tendex and vortex lines},''
{Phys.~Rev.~D} {\bf84},044037 ({\tt arXiv:1107.2959}) (2011).


\bibitem[{\citenamefont{{Schnittman} et~al.}(2008)\citenamefont{{Schnittman},
  {Buonanno}, {van Meter}, {Baker}, {Boggs}, {Centrella}, {Kelly}, and
  {McWilliams}}}]{SchnittmanEtAl}
\bibinfo{author}{\bibfnamefont{J.~D.} \bibnamefont{{Schnittman}}},
  \bibinfo{author}{\bibfnamefont{A.}~\bibnamefont{{Buonanno}}},
  \bibinfo{author}{\bibfnamefont{J.~R.} \bibnamefont{{van Meter}}},
  \bibinfo{author}{\bibfnamefont{J.~G.} \bibnamefont{{Baker}}},
  \bibinfo{author}{\bibfnamefont{W.~D.} \bibnamefont{{Boggs}}},
  \bibinfo{author}{\bibfnamefont{J.}~\bibnamefont{{Centrella}}},
  \bibinfo{author}{\bibfnamefont{B.~J.} \bibnamefont{{Kelly}}},
  \bibnamefont{and} \bibinfo{author}{\bibfnamefont{S.~T.}
  \bibnamefont{{McWilliams}}}, \bibinfo{journal}{\prd}
  \textbf{\bibinfo{volume}{77}}, \bibinfo{pages}{044031}
  (\bibinfo{year}{2008}), \eprint{arXiv:0707.0301}.



\bibitem[{\citenamefont{{Sundararajan}
  et~al.}(2010)\citenamefont{{Sundararajan}, {Khanna}, and
  {Hughes}}}]{SundararajanEtAlI}
\bibinfo{author}{\bibfnamefont{P.~A.} \bibnamefont{{Sundararajan}}},
  \bibinfo{author}{\bibfnamefont{G.}~\bibnamefont{{Khanna}}}, \bibnamefont{and}
  \bibinfo{author}{\bibfnamefont{S.~A.} \bibnamefont{{Hughes}}},
  \bibinfo{journal}{\prd} \textbf{\bibinfo{volume}{81}},
  \bibinfo{pages}{104009} (\bibinfo{year}{2010}), \eprint{arXiv:1003.0485}.


\bibitem{paper1} R.~H.~Price, G.~Khanna, and S.~A.~Hughes, {\prd}
 {\bf83}, 124002 (2011),
\eprint{arXiv:gr-qc/1104.0387}, ``Paper 1.''

\bibitem{BL}
R.~H.~Boyer and R.~W.~Lindquist, J.~Math.~Phys. {\bf8}, 265-281 (1967).


\bibitem{OT}A.~Ori and K.~S.~Thorne, Phys.\ Rev.~D {\bf62},  124022 (2000).

\bibitem{BD}
A.~Buonanno and T.~Damour, Phys.\ Rev.\ D {\bf62},  064015 (2000).


\bibitem{BPT}J.~M.~Bardeen, W.~H.~Press and S.~A.~Teukolsky,
  Astrophys.\ J.\ {\bf178}, 347 (1972).




\end{thebibliography}
\end{document}